\documentclass{amsart}
\usepackage{comment,amssymb,latexsym,cite,upref}
\usepackage{finalT}

\numberwithin{equation}{section}

\input cpne.def

\begin{document}

\title[Cosmological post-Newtonian expansions]{Cosmological post-Newtonian expansions to arbitrary order}
\author[T.A. Oliynyk]{Todd A. Oliynyk}
\address{School of Mathematical Sciences\\
Monash University, VIC 3800\\
Australia}
\email{todd.oliynyk@sci.monash.edu.au}

\begin{abstract}
We prove the existence of a large class of
one parameter families of solutions
to the Einstein-Euler equations that depend on the singular
parameter $\ep=v_T/c$ $(0<\ep < \ep_0)$, where $c$ is the speed of light, and $v_T$ is
a typical speed of the gravitating fluid. These solutions are shown to exist on a common
spacetime slab $M\cong [0,T)\times \Tbb^3$, and converge as $\ep \searrow 0$ to a solution
of the cosmological Poisson-Euler equations of Newtonian gravity. Moreover, we establish that these
solutions can be expanded in the parameter $\ep$
to any specified order with expansion coefficients that satisfy $\ep$-independent (nonlocal)
symmetric hyperbolic equations.
\end{abstract}

\maketitle 
\sect{intro}{Introduction}

Einstein's general relativity is presently the
most accurate theory of gravity. To completely determine
the gravitational field, the Einstein field equations
must be solved. These equations are extremely complex and outside of a small set of
idealized situations, they are impossible to solve
directly. However, to make physical predictions or understand
physical phenomena, it is often enough to find approximate
solutions that are governed by a simpler set of equations. The prime example
of this is Newtonian gravity which approximates general relativity very
well in regimes where the typical velocity of the gravitating
matter is small compared to the speed of light. Indeed, Newtonian
gravity successfully explains much of the behavior of our solar
system and is a simpler theory of gravity that is less difficult
to solve. By generalizing Newtonian gravity to the cosmological setting \cite{RS97},
it appears that the Newtonian theory can accurately describe gravity on
all scales except in regions near compact neutron
stars or black holes \cite{IW06,HN06}.

Although Newtonian gravity is quite
accurate, there are many situations where more accuracy is required and general relativistic effects must be
included. The \emph{post-Newtonian expansions} were developed as a method to include relativistic corrections to Newtonian gravity such
as gravitational lensing effects, and energy loss through gravitational radiation.
The goal of the post-Newtonian expansions is to
approximate
solutions to the Einstein field equations by a series
expansion in the small parameter
\eqn{epdef}{
\ep = \frac{v_T}{c}
}
where $c$ is the speed of light and $v_T$ is a typical speed associated with the
gravitating matter. A number of formal calculational schemes for computing the post-Newtonian
expansions have been developed and are widely used in both the asymptotically flat and cosmological
settings \cite{Blan,Chug,Fut96,FI,MT,HNP08,SA95,TF99}.
These post-Newtonian computational schemes are one of the most important
techniques in general relativity for calculating physical
quantities for the purpose of comparing with experiment. Implicitly, they
rely on the assumption that there actually exists solutions to the Einstein equations
that admit post-Newtonian expansions to a certain order. This leaves open the question of
existence of the post-Newtonian expansions. In view of the importance of the post-Newtonian
expansions, this is a question of considerable interest.

The main difficulty in determining when, and in what sense, the post-Newtonian expansions
approximate a relativistic solution is that the region of validity for
the post-Newtonian expansions is where $\ep=v_T/c$ is close to zero. Therefore to
understand the post-Newtonian expansions, solutions
of general relativity must be examined in the limit that $\ep \searrow 0$. In this limit the Einstein
field equations become \emph{singular} as the field equations contain
terms of the form $1/\ep$ that become unbounded as $\ep \searrow 0$.
The first person to overcome this difficulty and establish the existence of the
$0^{\text{th}}$ order post-Newtonian expansion (i.e. the Newtonian limit)
without any special assumptions such as spherical symmetry was Rendall in \cite{Ren94}.
In this paper, Rendall established the existence of
a large class of one-parameter families of solutions to the Einstein-Vlasov equations that
converge as $\ep \searrow 0$ to solutions of the Poisson-Vlasov equations of Newtonian gravity.
Using a different method based on techniques pioneered and developed by Klainerman, Majda, Kreiss,
and Schochet \cite{KM82,Kreiss80,Scho86,Scho88} to study singular limits of first order symmetric hyperbolic systems, we established
a similar result for the Einstein-Euler equations \cite{Oli06} and subsequently generalized
it to prove the existence of solutions to the Einstein-Euler equations that admit
post-Newtonian expansions to the $1^{\text{st}}$ order \cite{Oli07}.  We also note
that an improvement to at least the $2^{\text{nd}}$ post-Newtonian order is
possible using the results of \cite{Oli08}.

In this article, we adapt the methods of \cite{Oli06,Oli07} to the cosmological setting. We recall that the
Einstein-Euler equations, which govern gravitating
perfect fluids, are given by
\eqn{EEeqnA}{
G^{ij} = \frac{8\pi G}{c^4} T^{ij}\,
\AND \nabla_{i} T^{ij} = 0,} where \eqn{EEdefsA}{ T^{ij} = (\rho + c^{-2}p)v^i v^j + pg^{ij},  } with $\rho$ the fluid density,
$p$ the fluid pressure, $v^i$ the fluid
four-velocity normalized
by $v^i v_i = -c^2$, $c$ the speed of light, and $G$ the
Newtonian gravitational constant. As shown in \cite{Oli06}, these equations,
upon suitable rescaling, can be written
in the form \leqn{EEeqn}{ G^{ij} = 2\ep^4 T^{ij} \AND
\nabla_{i} T^{ij} = 0, } where \leqn{EEdefs}{ T^{ij} = (\rho+\ep^2 p)v^i v^j + p g^{ij} \AND v^i v_i = -\frac{1}{\ep^2}. }
In this formulation, the fluid four-velocity
$v^i$, fluid density $\rho$, fluid pressure $p$, the
metric $g_{ij}$, and the coordinates $(x^i)$ $i=0,\ldots,3$ are
dimensionless. In this article, we restrict ourselves to cosmological spacetimes of the form\footnote{The results of this article
 can be extended to other interesting spacetimes. A general discussion of spacetimes that admit Newtonian limits or post-Newtonian expansions
 will be presented in an upcoming paper.}
 $M=[0,T)\times \Tbb^3$, and we will always use $x=(x^I)$ $I=1,2,3$ to denote the standard periodic coordinates (with period 1) on
the torus $\Tbb^3= S^1\times S^1 \times S^1$. The coordinate $x^0$ will parametrize the interval $[0,T)$ and  $t=x^0/v_T$ will denote an absolute Newtonian time coordinate. By a choice of units,
we can and shall set $v_T=1$, in which case $t=x^0$.

The main aim of this article is to present a proof that establishes the existence of a large class of one-parameter family of
solutions $(g^\ep_{ij},v_\ep^i,\rho_\ep,p_\ep)$  $0<\ep<\ep_0$
to the Einstein-Euler equations \eqref{EEeqn} that \textbf{(i)} exist on a common
spacetime slab $M=\Tbb^3\times [0,T)$, and
\textbf{(ii)} can be expanded in the parameter $\ep$ to any finite order  $\ell \in \Zbb_{\geq 0}$ with expansion coefficients that are $\ep$-independent. For a precise version of this statement, see Theorem \ref{mthm} below.
To agree with standard terminology, we will say that these solutions admit a
\emph{$(\ell/2)^{\text{th}}$ order (cosmological) post-Newtonian expansion}.

In light of the significant and well-known difficulties that are encountered at both the formal and rigorous level in trying to develop
post-Newtonian expansions on asymptotically flat spacetimes beyond the order $2.5$ \cite{Ren92a}, it is somewhat surprising that
these difficulties are absent in the cosmological setting. On asymptotically flat spacetimes, the problems that
occur in the higher order post-Newtonian expansions are often attributed to the reaction of gravitational
radiation with itself and matter. The analysis contained in this paper shows that this is not the complete
story as the these effects are also present in the cosmological setting, but do not cause similar difficulties\footnote{We will address this discrepancy
more thoroughly in a separate article.}.

In the cosmological setting, the simplest one-parameter
family of solutions that admits a post-Newtonian expansion to any order\footnote{This is an exceptional solution. Generically, it will only be possible to expand
dynamical solutions to a fixed finite order in $\ep$. } is the FLRW dust solution given by $(g^\ep_{ij}=h_{ij},v^i_\ep=\xi^i,\rho_\ep = \mu,p_\ep=0)$
where
\lalign{flrw1}{
h_{ij} & = -\frac{1}{\ep^2}\delta_i^0\delta_j^0
+ a(t)\delta_i^I\delta_j^J\delta_{IJ} , \label{flrw1.1} \\
\xi^i & = \delta^i_0, \label{flrw1.2} \\
\mu(t) & = \frac{3}{8} \left[\frac{a'(t)}{a(t)}\right]^2, \label{flrw1.3} \\
}
and $a(t)$ satisfies the equation
\leqn{flrw2}{
\frac{a''(t)}{a(t)} -\frac{1}{2}\left[\frac{a'(t)}{a(t)}\right]^2 + \frac{2}{3}\mu(t) = 0.
}
As is well known, the differential equation \eqref{flrw2} can be integrated explicitly to give
\leqn{flrw3}{
a(t) = a_0\left(\frac{2}{3\mu_0}\right)^{-\frac{2}{3}}\left(t+\sqrt{\frac{2}{3\mu_0}}\right)^{\frac{4}{3}}, \AND
\mu(t) = \frac{2}{3}\left(t+\sqrt{\frac{2}{3\mu_0}}\right)^{-2},
}
where $a_0=a(0)$ and $\mu_0=\mu(0)$ are positive constants\footnote{With these choices,
the big bang occurs at ``time'' $t=-\sqrt{\frac{2}{3\mu_0}}$}.

In this article, we will, for simplicity, always assume an adiabatic equation of state of the form
\leqn{eos}{
p = f(\rho),
}
where $f: \Rbb_{>0} \rightarrow \Rbb_{>0}$ is an analytic function that satisfies
$f'(\rho) > 0$ for all $\rho > 0$. Even though the FLRW dust solution \eqref{flrw1.1}-\eqref{flrw3}
does not arise from an equation of state of the form \eqref{eos}, it plays the role of defining
a \emph{Newtonian background} that is essential for analyzing the limit $\ep \searrow 0$ and generating the $\ep$-expansion.
The role of the FLRW solution is easiest to see at the $0^{\text{th}}$ order where it is used to define the \emph{cosmological Poisson-Euler
equations}
\lalign{PE}{
\del_t \oset{0}{\rho} + \del_I\bigl(\oset{0}{w}{}^I \oset{0}{\rho}\bigr) &= -\frac{3}{2}\frac{a'(t)}{a(t)} \oset{0}{\rho}
 && (\del_I = \del_{x^I}),\label{PE.1} \\
\oset{0}{\rho}\bigl(\del_t \oset{0}{w}{}^J + \oset{0}{w}{}^I\del_I \oset{0}{w}{}^{J}\bigr) &= -\frac{1}{a(t)}
\bigl(\del^J f(\oset{0}{\rho}) + \oset{0}{\rho}\del^J \oset{0}{\Phi} + a'(t)\oset{0}{\rho}\oset{0}{w}{}^J\bigr)
&& (\del^I = \delta^{IJ}\del_J), \label{PE.2}\\
\Delta \oset{0}{\Phi} & = a(t) \left(\oset{0}{\rho} - \int_{\Tbb^3}\oset{0}{\rho} \, d^3 x\right) && (\Delta = \delta^{IJ}\del_I\del_J), \label{PE.3}
}
of (cosmological) Newtonian gravity. We note that these equations agree with the Newton-Cartan field equations
for a gravitating fluid formulated in adapted coordinates \cite{Kunz76,RS97}.

The cosmological Poisson Euler equations can be brought into a more familiar form by introducing Galilei coordinates \cite{Kunz76,RS97}. This is
done as follows: suppose $\{\oset{0}{\rho}(t,x),\oset{0}{w}{}^I(t,x),\oset{0}{\Phi}(t,x)\}$ is a solution of the cosmological
Poisson-Euler equations \eqref{PE.1}-\eqref{PE.3} on $M = [0,T)\times \Tbb^3$. Then, letting
$\tilde{M} = [0,T)\times \Rbb^3$ denote the covering space, we define a diffeomorphism on $\tilde{M}$ by
\eqn{phihdef}{
\psi \: : \: \tilde{M} \longrightarrow \tilde{M} \; : \: (t,x) \longmapsto (t,x/\sqrt{a(t)}).
}
Lifting the cosmological Poisson-Euler equations to $\tilde{M}$, and then pulling back by $\psi$ shows that\footnote{In the Newton-Cartan theory, the
fluid velocity 3-vector $\oset{0}{w}{}^I$ is the spatial part of a 4-vector $\oset{0}{w} = \del_t + \oset{0}{w}{}^I\del_I$ \cite{RS97}. The formula
\eqref{hatvars1.2} follows from the calculating the spatial components of $\hat{w} = \psi^* \oset{0}{w}$. The other two formulas
\eqref{hatvars1.1} and \eqref{hatvars1.3} just follow from the definition of the pullback, i.e.
$\hat{\rho} = \psi^*\oset{0}{\rho}$ and $\hat{\Phi} = \psi^*\oset{0}{\Phi}$.}
\lalign{hatvars1}{
\hat{\rho}(t,x) &= \oset{0}{\rho}\bigl(t,x/\sqrt{a(t)}\bigr), \label{hatvars1.1} \\
\hat{w}^J(t,x) & = \sqrt{a(t)}\oset{0}{w}{}^J\bigl(t,x/\sqrt{a(t)}\bigr)+ \frac{1}{2}\frac{a'(t)}{a(t)} x^J, \label{hatvars1.2}\\
\hat{\Phi}(t,x) &= \oset{0}{\Phi}\bigl(t,x/\sqrt{a(t)}\bigr), \label{hatvars1.3}
}
satisfy
\lalign{limC}{
\del_t \hat{\rho} & = -\hat{w}^I\del_I \hat{\rho} - \hat{\rho}\del_I \hat{w}^I, \label{limC.1} \\
\del_t \hat{w}^J & = -\hat{w}^I\del_I \hat{w}^J - \frac{1}{\hat{\rho}}\del^J f(\hat{\rho}) + \hat{g}^J, \label{limC.2} \\
\Delta \hat{\Phi} & = 4\left(\hat{\rho} - \int_{\Tbb^3}\oset{0}{\rho} \, d^3 x\right), \label{limC.3}
}
where
\eqn{hatvars2}{
\hat{g}^J =
- \frac{1}{4}\del^J\Phih -\frac{1}{3} \left(\int_{\Tbb^3}\oset{0}{\rho} \, d^3 x \right)  x^J .
}
Defining a Newtonian potential by
\eqn{hatvars3}{
\check{\Phi} = \frac{\hat{\Phi}}{4} + \frac{1}{6}\left( \int_{\Tbb^3}\oset{0}{\rho} \, d^3 x\right) \delta_{IJ}x^I x^J,
}
a short calculation shows that $\check{\Phi}$ satisfies the Poisson equation
\leqn{hatvars4}{
\Delta \check{\Phi} = \hat{\rho}
}
while the acceleration due to gravity $\hat{g}^J$ takes the familiar
form
\leqn{hatvars5}{
\hat{g}^J = - \del^J \check{\Phi}.
}
Together, equations \eqref{limC.1}, \eqref{limC.2}, \eqref{hatvars4}, and \eqref{hatvars5} demonstrate that
solutions to the cosmological Poisson-Euler equations determine solutions to the standard Poisson-Euler
equations on the covering space $\tilde{M}$.

To understand the limit $\ep \searrow 0$, we use a slight variation of the approach using in \cite{Oli06} and  replace
the metric $g_{ij}$ and the fluid velocity
$v^i$ with new variables that are compatible with the limit
$\ep \searrow 0$. The new gravitational variable $\ub^{ij}$ is defined by
\leqn{metrecA}{
g^{ij} = \frac{q^{ij}}{\sqrt{-|h|\det(q^{kl})}} \qquad (|h| = -\det (h_{ij}) ),
}
where
\leqn{metrecB}{
q^{ij} = h^{ij} + \ep^2 J^i_k J^j_l \ub^{kl}, \AND J^j_i = \ep \delta^j_0\delta^0_i + \delta^j_I \delta^I_i,
}
while the new fluid four-velocity $w^i$ is defined by
\leqn{wdef.intro}{
v^i = (1 + \ep w^0) \delta^i_0 + \delta^i_I w^I.
}
From these formulas it is not difficult to see that $\ub^{ij}$ and $w^i$ are equivalent
to the metric $g_{ij}$ and the fluid four-velocity $v^{i}$, respectively, for $\ep>0$ and
are well defined at $\ep=0$.

Following \cite{Oli06,Oli07}, these variables combined with a harmonic gauge can be used
to cast the Einstein-Euler equations into
a singular  (nonlocal) symmetric hyperbolic system of the form
\eqn{EFsym2.intro}{ A^0(t,\ep,W)\partial_t W =
\frac{1}{\epsilon}C^I\partial_I W + A^I(t,\ep,W)\partial_I W +
F(t,\ep,W). }
This equation has the necessary structure that is required to use the results of
\cite{KM82,Kreiss80,Scho86} to study the limit $\ep \searrow 0$, and to use Kreiss' bounded derivative principle \cite{BK,Kreiss91,Scho88}to generate
$\ep$-expansions. The beauty of Kreiss' bounded derivative principle is that it
reduces the
problem of generating $\ell^{\text{th}}$ order expansions in $\ep$ to that of constructing initial data
for which the time derivatives $(\del_t^p W)|_{t=0}$ are bounded as $\ep\searrow 0$ for $p=1,2,\ldots, \ell+1$.
This process of
choosing such initial data is called \emph{initialization}. Thus the existence problem for the post-Newtonian expansions is
replaced by the problem of establishing the existence of initial data that can be properly initialized.  The main technical result of
this paper is to construct a large class of initial data for the Einstein-Euler equations that can be initialized to any order.
The method for constructing this data is based on a technique introduced by Lottermoser in \cite{Lott} who was the first
person to prove the existence of a one-parameter family of initial data that depends analytically
on $\ep$ and converges to the expected Newtonian initial data as $\ep \searrow 0$.

It is worthwhile to note that initializing the data to arbitrary order does not seem to be possible in
the asymptotically flat case although it is possible for low orders \cite{Oli07,ILR}. In any case,
initializing the data, whenever possible, provides a constructive method for generating
initial data that has minimal initial gravitational radiation to an accuracy that is governed
by the order of the $\ep$-expansion. This is certainly a significant improvement over other ad-hoc methods
for choosing initial data that have been previously used in the literature.

\subsect{not}{Notation} Before stating the main result of this article, we first introduce a number of
function spaces. Given a finite dimensional vector space $V$, we let $H^s(V)$ denote the standard  Sobolev space of $V$-valued
maps on $\Tbb^3$. When $V=\Rbb$, we just write $H^s$. The only two vector spaces that will be used in
this article are  $\Rbb^N$ and the space of symmetric matrices $\Sbb{N} = \{ \: (u^{ij})\in \Mbb{N}\: |\: u^{ij} = u^{ji} \: \}$.
Letting
\eqn{L2ip}{
\ip{\psi_1}{\psi_2}_{L^2} = \int_{\Tbb^3}\psi_1 \psi_2 d^3x = \int_{0}^1\int_0^1 \int_0^1 \psi_1(x^1,x^2,x^3) \psi_2(x^1,x^2,x^3) dx^1 dx^2 dx^3
}
denote the standard $L^2$ inner product, we denote the projection operator
onto the $L^2$ orthogonal complement of the constant function $1$ by
\leqn{L2proj}{
\Pi(\psi) = \psi - \ip{1}{\psi}_{L^2}1 \quad \forall \; \psi\in L^2(\Tbb^3).
}
Given $\{\ev_{\alpha}\}_{\alpha=1}^N$ any basis for $V$, we use the projection \eqref{L2proj} to define
\eqn{Hsbar}{
\Ho^s(V)= \Bigl\{\, \psi(x^I)=\sum_{\alpha=1}^N \psi^\alpha(x^I)\ev_\alpha \in H^s(V) \: \Bigl|\:  \ip{1}{\psi^\alpha}_{L^2} = 0 \quad \text{for $\alpha=1,2,\ldots,N$}\: \Bigr\}.
}
We also define the standard hyperbolic evolution spaces
\eqn{Xspace}{
X_{T,\ell,s}(V) = \bigcap_{p=0}^{\ell+1} C^p\bigl([0,T), H^{s-p}(V)\bigr).
}
and write $X_{T,\ell,s}$ if $V=\Rbb$.
Finally, given two Banach spaces $X,Y$ and a  open set $U\subset X$, we let
$C^\omega(U,Y)$ denote the set of analytic maps $f: U \rightarrow Y$.

\subsect{mres}{Main Theorem}

The following Theorem contains the precise statement of the existence of
post-Newtonian expansions to arbitrary orders. The proof of the Theorem
can be found in Section \ref{proof}.
\begin{thm} \label{mthm} \mnote{[mthm]}
Suppose $\ell \in \Zbb_{\geq 0}$, $s > 3/2+\ell+3$, $C_s$ is the constant appearing in the inequality \eqref{BA},
$\delta = \mu_0/(2C_s)$, $y_0^{ij} \in \Sbb{4}$, $y_1^{IJ}\in \Sbb{3}$,
$\rhot_0 \in B_\delta(\Ho^s)$, $\wt^I_0\in \Ho^{s}(\Rbb^3)$, $\ut^{IJ}_\ell \in \Ho^{s+1-\ell}(\Sbb{3})$,
$\ut^{IJ}_{\ell+1} \in \Ho^{s-\ell}(\Sbb{3})$, and let $T_0$ be the maximal time of existence\footnote{See Proposition \ref{limA}} of solutions of the
Poisson-Euler equations \eqref{PE.1}-\eqref{PE.3} with initial data $\oset{0}{\rho}|_{t=0} = \mu_0 + \rhot_0$ and
$\oset{0}{w}{}^I|_{t=0} = \wt_0^I/(\mu_0+\rhot_0)$. Then for any $T<T_0$
there exists an $\ep_0 >0$, and maps
\alin{mth1}{
&\ub_\ep^{ij} \quad : \quad \ub_\ep^{ij},\; \del_I \ub^{ij}_\ep,\; \del_t\ub^{ij}_\ep \in X_{T,\ell+2,s}(\Sbb{4})
 \quad 0<\ep < \ep_0, \\
&\rho_\ep \in X_{T,\ell+2,s}, \quad w^i_\ep \in X_{T,\ell+2,s}(\Rbb^4)  \quad  0 < \ep < \ep_0 ,\\
& \oset{0}{\rho} \in X_{T_0,\ell+2,s}, \quad \oset{0}{w}{}^I \in X_{T_0,\ell+2,s}(\Rbb^3), \quad  \oset{0}{\Phi} \in
X_{T_0,\ell+2,s+2},\\
& \oset{p}{\ub}{}^{ij} \quad :
\quad \oset{p}{\ub}{}^{ij},\;
\del_I\oset{p}{\ub}{}^{ij}, \; \del_t\oset{p}{\ub}{}^{ij} \in X_{T,\ell+2-p,s-p}(\Sbb{4}) \quad
p=1,2,\ldots \ell, \\
& \oset{p}{\rho} \in X_{T,\ell+2-p,s-p}, \quad \oset{p}{w}{}^i \in
X_{T,\ell+2-p,s-p}(\Rbb^4) \quad p=1,2,\ldots \ell, \\
& \oset{p}{\ub}{}_\ep^{ij} \quad : \quad
 \oset{p}{\ub}{}_\ep^{ij},\;
\del_I\oset{p}{\ub}{}_\ep^{ij},\; \del_t \oset{p}{\ub}{}_\ep^{ij} \in X_{T,1,s-\ell-1}(\Sbb{4})\quad
(p,\ep)\in \mathbb{Z}_{\geq \ell+1}\times (0,\ep_0),\\
& \oset{p}{\rho}_\ep \in  X_{T,1,s-\ell-1}  , \quad \oset{p}{w}{}_\ep^i \in
X_{T,1,s-\ell-1}(\Rbb^4) \quad (q,\ep)\in \mathbb{Z}_{\geq \ell+1}\times (0,\ep_0), \\
& \lambda_0 \in C^\omega((-\ep_0,\ep_0),\Rbb), \quad z^I_0 \in C^\omega((-\ep_0,\ep_0),\Rbb^3), \quad \lambda_0(0)= z^I_0(0)=0,
}
such that
\begin{itemize}
\item[(i)] the triple $\{\ub^{ij}_\ep,\rhot_\ep,w^i_\ep\}$ determines, via
formulas \eqref{metrecA}-\eqref{wdef.intro},  a unique solution
to the Einstein-Euler equations \eqref{EEeqn} in the harmonic gauge for $0<\ep < \ep_0$ on
the spacetime region $(t=x^0,x^I) \in M=[0,T)\times \Tbb^3$,
\item[(ii)] \gath{idatdef}{
\ip{1}{\ub_\ep^{ij}|_{t=0}}_{L^2}= y^{ij}_0, \quad \ip{1}{\del_t\ub_\ep^{ij}|_{t=0}}_{L^2}=y^{IJ}_1, \\
\Pi\bigl(\del_t^{\ell+1} \ub_\ep^{IJ}|_{t=0}\bigr) = \ep^2 \ut^{IJ}_{\ell+1}, \quad  \Pi\bigl(\del_t^{\ell+2} \ub_\ep^{IJ}|_{t=0}\bigr) = \ep^2 \ut^{IJ}_{\ell+2}, \\
\rho_\ep|_{t=0}= \mu_0 + \lambda_0(\ep) + \rhot_0, \AND w^I_\ep|_{t=0} = (z_0^I(\ep) + \wt_0^I)/(\rho_\ep|_{t=0})
} for $0<\ep  < \ep_0$,
\item[(iii)] $\{\oset{0}{\rho},\oset{0}{w}{}^I,\oset{0}{\Phi}\}$ is the unique solution
to the Poisson-Euler equations \eqref{PE.1}-\eqref{PE.3} with
initial data $\oset{0}{\rho}|_{t=0} = \mu_0 + \rhot$ and
$\oset{0}{w}{}^I|_{t=0} = \wt_0^I/(\mu_0+\rhot)$,
\item[(iv)] for $p=1,2,\ldots,\ell$, $\{\oset{p}{\ub}{}^{ij},\oset{p}{\rho},\oset{p}{w}{}^i\}$ satisfies a linear (nonlocal) symmetric hyperbolic system
that only depends on
and $\{\oset{0}{\rho},\oset{0}{w}{}^I,\oset{0}{\Phi}, \oset{q}{\ub}{}^{ij},\oset{q}{\rho},\oset{q}{w}{}^i \: |\:
q=1,2,\ldots,p-1  \}$,
\item[(v)] for $p\in \Zbb_{\geq \ell+1}$,  $\{\oset{p}{\ub}{}_\ep^{ij},
\oset{p}{\rho}_\ep,\oset{p}{w}{}_\ep^i\}$ satisfies a linear (nonlocal) symmetric hyperbolic system
that only depends on $\ep$, $\{\oset{0}{\rho},\oset{0}{w}{}^I,\oset{0}{\Phi}, \oset{q}{\ub}{}^{ij},\oset{q}{\rho},\oset{q}{w}{}^i \: |\:
q=1,2,\ldots,\ell  \}$ and
 $\{\oset{q}{\ub}_\ep{}^{ij},\oset{q}{\rho}_\ep,
\oset{q}{w}{}_\ep^i \: | \: q=\ell+1,\ell+2,\ldots,p-1 \}$,
\item[(vi)]
 $\{\ub^{ij}_\ep,\rho_\ep,w^i_\ep\}$
and $\{\oset{p}{\ub}{}_\ep^{ij},
\oset{p}{\rho}_\ep,\oset{p}{w}{}_\ep^i\}$ $(p\geq \ell+1)$
satisfy the estimates:
\alin{mth2}{
&
 \norm{\ub_\ep^{ij}(t)}_{H^{s+1}} +\ep\norm{\del_t\ub_\ep^{ij}(t)}_{H^s}
+\ep\norm{\del_t\del_I\ub_\ep^{ij}(t)}_{H^{s-1}}
+\ep^2\norm{\del_t^2\ub_\ep^{ij}(t)}_{H^{s-1}} \lesssim 1,\\
& \norm{\rho_\ep(t)}_{H^{s}}+\norm{w^i_\ep(t)}_{H^{s}}
+ \norm{\del_t\rho_\ep(t)}_{H^{s-1}} + \norm{\del_t w^i_\ep(t)}_{H^{s-1}} \lesssim 1, \\
 & \norm{\oset{p}{\ub}{}_\ep^{ij}(t)}_{H^{s-\ell}} +
\ep\norm{\del_t\oset{p}{\ufb}{}_\ep^{ij}(t)}_{H^{s-\ell-1}}
+\ep\norm{\del_t\del_I\oset{p}{\ub}{}_\ep^{ij}(t)}_{H^{s-\ell-2}}
+\ep^2\norm{\del_t^2\oset{p}{\ub}{}_\ep^{ij}(t)}_{H^{s-\ell-2}} \lesssim 1,\\
& \norm{\oset{p}{\rho}_\ep(t)}_{H^{s-\ell-1}}+\norm{\oset{p}{w}{}_\ep^i(t)}_{H^{s-\ell-1}}
+ \norm{\del_t\oset{p}{\rho}_\ep(t)}_{H^{s-\ell-2}}+ \norm{\del_t
\oset{p}{w}{}_\ep^i(t)}_{H^{s-\ell-2}} \lesssim 1,
}
for all $(t,\ep) \in [0,T_0)\times (0,\ep_0)$, and
\item[(vii)]  $\{\ub^{ij},\rho_\ep,w^i_\ep\}$ admits
convergent expansions (uniform for $0 < \ep < \ep_0$) of the form
\alin{mth3}{
\ub^{ij}_\ep&= 4\delta^i_0\delta^j_0\oset{0}{\Phi} + \sum_{p=1}^\ell
\ep^p \oset{p}{\ub}{}^{ij} +
\sum_{p=\ell+1}^\infty \ep^p \oset{p}{\ub}{}_\ep^{ij}, \\
\ep^\nu\del_t^\nu\del_I\ub^{ij}_\ep &= 4\ep^\nu
\delta^i_0\delta^j_0\del_t^\nu\del_I\oset{0}{\Phi} + \sum_{p=1}^\ell
\ep^{p+\nu} \del_t^\nu\del_I\oset{p}{\ub}{}^{ij} +
\sum_{p=\ell+1}^\infty \ep^{p+\nu} \del_t^\nu\del_I\oset{p}{\ub}{}_\ep^{ij},  \\
\ep^{\nu+1} \del_t^{\nu+1} \ub^{ij}_\ep &= 4\ep^{\nu+1} \delta^i_0\delta^j_0
\del_t^{\nu+1} \oset{0}{\Phi} + \sum_{p=1}^\ell
\ep^{p+\nu+1} \del_t^{\nu+1} \oset{p}{\ub}{}^{ij} +
\sum_{p=\ell+1}^\infty \ep^{p+\nu+1} \del_t^{\nu+1} \oset{p}{\ub}{}_\ep^{ij}, \\
\del_t^\nu \rho_\ep & =
\del_t^\nu \oset{0}{\rho} + \sum_{p=1}^\ell \ep^p\del_t^\nu \oset{p}{\rho}
+\sum_{p=\ell+1}^\infty \ep^p\del_t^\nu \oset{p}{\rho}_\ep,\\
\del_t^\nu w^i_\ep & =
\del_t^\nu \oset{0}{w}{}^i + \sum_{p=1}^\ell \ep^p\del_t^\nu \oset{p}{w}{}^i
+\sum_{p=\ell+1}^\infty \ep^p\del_t^\nu \oset{p}{w}{}_\ep^i ,
}
where the expansions are convergent in $C^0([0,T_0);H^{s-\ell-2-\nu})$ for $\nu=0,1$.
\end{itemize}
\end{thm}
An important point not explicitly stated in the above Theorem but follows from the proof
is that for $p=1,2,\ldots,\ell$ the equations satisfied by coefficients
$\{\oset{p}{\ub}{}^{ij},\oset{p}{\rho},\oset{p}{w}{}^i\}$
from Theorem \ref{mthm} (iv)
can be derived by assuming a harmonic gauge \eqref{harm1} and
substituting the expansions of Theorem \ref{mthm} (vii) in the Einstein-Euler equations
\eqref{EEeqn} and collecting terms together of the same order in $\ep$ up to order $\ell$.
It is this fact that guarantees that the expansions of Theorem \ref{mthm} (vii) coincide
with the post-Newtonian expansions of order $\ell/2$.

As a final remark, all of the results of this article can be adapted to include a cosmological
constant $\Lambda$. The basic changes needed to be made include replacing the stress
energy tensor \eqref{EEdefs} by
\eqn{stressLambda}{ T^{ij} = (\rho+\ep^2 p)v^i v^j + p g^{ij} - \frac{\Lambda}{\ep^2}g^{ij},}
and replacing the FLRW equations \eqref{flrw1.3} and \eqref{flrw2} by
\gath{FLRWLambda}{
\mu(t) + \Lambda  = \frac{3}{8} \left[\frac{a'(t)}{a(t)}\right]^2,
\intertext{and}
\frac{a''(t)}{a(t)} -\frac{1}{2}\left[\frac{a'(t)}{a(t)}\right]^2 + \frac{2}{3}\bigl(\mu(t)+\Lambda\bigr) = 2\Lambda.
}
As shown in \cite{BRR}, the inclusion of a positive cosmological constant guarantees the long time existence
of small perturbations of the constant density solution $\{\oset{0}{\rho}=\mu,\oset{0}{w}{}^I=0,\oset{0}{\Phi}=0\}$
of the cosmological Poisson-Euler equations \eqref{PE.1}-\eqref{PE.3}. The importance of the long time existence
of solutions is that it is a necessary ingredient of any analysis of a lower
bound on the time of validity of the post-Newtonian expansions as a function of $\ep$. We plan to address this
problem of determining a lower bound on the time of existence of the post-Newtonian expansions in the near future.

\sect{edeqns}{The Einstein-Euler equations}

\subsect{red}{Reduced Einstein Equations}

In order to derive a suitable symmetric hyperbolic system for
the gravitational field equations, we introduce new
coordinates related to old ones by the rescaling
\eqn{bcoords}{ \xb^0 = x^0/\epsilon, \quad \xb^J = x^J,} and let
\eqn{bpartial}{
\quad \delb_i =  \frac{\partial\;}{\partial \xb^i} \, . }
In
the new coordinates, the spacetime metric $\gb_{ij}$ and the FLRW metric $\hb_{ij}$ (see \eqref{flrw1.1})
are given by \eqn{bmetricdef}{ \gb_{ij} = J^k_i J^k_j g_{ij} \AND
\hb_{ij} = -\delta^0_i\delta^0_j + a(t)\delta^I_i\delta^J_i \delta_{IJ},
}
respectively, where $J^j_i$ is defined in \eqref{metrecB}.
For latter use, we record the non-zero independent components of the Christoffel symbols $\gammab^k_{ij}$ and the curvature
$\Rcb_{ijkl}$ of the metric $\hb_{ij}$:
\lalign{hcurv}{
\gammab^I_{0I} &= \frac{\ep}{2}\frac{a'(t)}{ a(t)}, \label{hcurv.1} \\
\gammab^0_{II} & = \frac{\ep}{2} a'(t), \label{hcurv.2} \\
\Rcb_{0I0I} &= -\frac{\ep^2}{4}\frac{2 a(t) a''(t) - [a'(t)]^2}{a(t)}, \label{hcurv.3}
\intertext{and}
\Rcb_{2121} &=\Rcb_{1313}=\Rcb_{2323} = \frac{\ep^2}{4}[a'(t)]^2.
}

As discussed in the introduction, we take the symmetric 2-tensor $\ub^{ij}$ as our primary gravitational variable
where
\leqn{ubdef}{
\ub^{ij} =\frac{1}{\ep^2} \left( \frac{\sqrt{|\gb|}}{\sqrt{|\hb|}}\gb^{ij}-\hb^{ij}\right) \Longleftrightarrow
\sqrt{|\gb|}\gb^{ij} = \sqrt{|\hb|} {}(\hb^{ij}+\ep^2\ub^{ij}),
}
and
\leqn{voldef}{
|\gb| = -\det(\gb_{ij}) \AND |\hb| = -\det(\hb_{ij}) = [a(t)]^3.
}
Observe that the metric can be recovered from the $\ub^{ij}$ by the formula
\leqn{utog}{
\gb^{ij} = \frac{1}{\sqrt{-|\hb|\det(\gt^{kl})}}\gt^{ij},
}
where
\leqn{gtdef}{
\gt^{ij} = \hb^{ij} + \ep^2 \ub^{ij}.
}
Substituting \eqref{utog} in to the standard formula for the Christoffel symbols
gives
\leqn{christ}{
\Gammab^k_{ij} = \gammab^k_{ij} +\ep^2\Bigl(-\gt_{l(i}\Db_{j)}\ub^{kl}+\Half\gt^{kl}\gt_{im}\gt_{jn}\Db_l\ub^{mn}
-\Quarter\gt^{kl}\gt_{ij}\gt_{mn}\Db_l\ub^{mn} +\Half\gt_{lm}\delta^k_{(i}\Db_{j)}\ub^{lm}\Bigr),
}
where $(\gt_{ij})  = (\gt^{ij})^{-1}$ and $\Db_k$ is the $\hb_{ij}$ covariant derivative.
Using this formula, the Einstein tensor $\Gb^{ij}$ of the metric $\gb_{ij}$
is given by
\leqn{Gb1}{
|\gb|\Gb^{ij} = \frac{\ep^2}{2}|\hb| \left[
\gt^{kl}\Db_k\Db_l \ub^{ij} + \ep^2\bigl(a_1^{ij}+ a_2^{ij}+ a_3^{ij}\bigr) + b^{ij} + \ep c_1^{ij} + \ep^2
c_2^{ij} +
4\ep^2 \Tcb^{ij}
\right],
}
where
\lalign{Gb2}{
a_1^{ij} & = \Half \bigl(\Half \gt_{k l} \gt_{mn} - \gt_{km}
\gt_{l n} \bigr) \bigl(\gt^{ip} \gt^{jq} - \Half \gt^{ij}
\gt^{pq} \bigr)\Db_{p} \ub^{k l}\Db_{q} \ub^{mn}
\label{Gb2.1} ,\\
a_2^{ij} & = 2\gt_{k l}\bigl( \gt^{n(i }
\Db_{m} \ub^{j) l}\Db_{n} \ub^{k m} - \Half
\gt^{ij}\Db_{m} \ub^{k n} \Db_{n} \ub^{m l}
- \gt^{mn}\Db_{m}  \ub^{ik}\Db_{n} \ub^{j l}\bigr)\label{Gb2.2}, \\
a_3^{ij} & = \Db_k\ufb^{ij}\Db_{l} \ub^{k l}-
\Db_k\ub^{i l}\Db_l \ub^{jk}\label{Gb2.3}, \\
b^{ij} & =  \gt^{ij}\Db_k\Db_l \ub^{k l} -
2\Db_{l}\Db_k \ub^{k(i} \gt^{j)l} \label{Gb2.4}, \\
c_1^{ij} & = -\bigl(\hb^{ij}\ep\ub^{kl} + \ep\ub^{ij}\hb^{kl}\bigr)\frac{1}{\ep^2}\Rcb_{kl}
+\frac{2}{\ep^2}\Rcb_{lkm}{}^{(i}\ep\ub^{j)k}\hb^{lm}+ \frac{2}{\ep^2}\Rcb_{lkm}{}^{(i}\hb^{j)k}\ep\ub^{lm}, \label{Gb2.5} \\
c_2^{ij} & = -\ep\ub^{ij}\ep\ub^{kl}\frac{1}{\ep^2}\Rcb_{kl}+\frac{2}{\ep^2}\Rcb_{lkm}{}^{(i}\ep\ub^{j)k}\ep \ub^{lm},\label{Gb2.6}
\intertext{and}
\Tcb^{ij} & = \mu \xib^{i}\xib^{j}, \qquad \xib^{i} = \frac{1}{\ep} \delta^{i}_0. \label{Gb2.7}
}

To fix the gauge, we set
\leqn{harm1}{
\Db_i\ub^{ij} = 0.
}
For $\ep>0$, it is clear from \eqref{ubdef} that this is equivalent to
\eqn{harm2}{
\Db_i\bigl(\sqrt{|\gb|}\gb^{ij}\bigr)=0,
}
and this is easily seen to be equivalent to the harmonic coordinate condition
\eqn{harm3}{
\gb^{ij}\bigl(\Gammab^k_{ij}-\gammab^k_{ij}\bigr) = 0.
}
Defining the reduced Einstein tensor $\Gb_R^{ij}$ by
\leqn{redGb}{
\Gb_R^{ij} = \frac{1}{\ep^2}\frac{|\gb|}{|\hb|}\Gb^{ij} - b^{ij} = \Half\bigl(\gt^{kl}\Db_k\Db_l \ub^{ij} + \ep^2\bigl(a_1^{ij}+ a_2^{ij}+ a_3^{ij}\bigr) + \ep c_1^{ij} + \ep^2 c_2^{ij} +
4\ep^2 \Tcb^{ij} \bigr) ,
}
the Einstein equation $\Gb^{ij}=2\ep^4 \Tb^{ij}$ in the gauge \eqref{harm1} becomes
\leqn{redeqns}{
\Gb_R^{ij} = 2\ep^2 \frac{|\gb|}{|\hb|}\Tb^{ij},
}
where
\leqn{stress}{
\Tb^{ij} = (\rho+\ep^2 p)\vb^i\vb^j + p \gb^{ij} \AND \vb^i\vb_j = -\frac{1}{\ep^2}.
}

To write the reduced Einstein equations \eqref{redeqns} in first order form, we introduce the variables
\leqn{udef}{
u^{ij} = \ep \ub^{ij} \AND u^{ij}_k = \Db_k \ub^{ij}.
}
With these variables, we have that
\eqn{comm1}{
\Db_k u_l^{ij} = \Db_l u_k^{ij} -\frac{2}{\ep}\Rcb_{klm}{}^{(i}u^{j)m},
}
or equivalently
\leqn{comm2}{
\delb_k u_l^{ij} = \delb_l u_k^{ij} + 2\gammab_{lm}^{(i}u_k^{j)m} -  2\gammab_{km}^{(i}u_l^{j)m}
- \frac{2}{\ep}\Rcb_{klm}{}^{(i}u^{j)m}.
}
In particular, this implies that
\leqn{comm3}{
\delb_0 u_I^{ij} = \delb_I u_0^{ij} + 2\gammab_{Im}^{(i}u_0^{j)m} -  2\gammab_{0m}^{(i}u_I^{j)m}
- \frac{2}{\ep}\Rcb_{0Im}{}^{(i}u^{j)m},
}
and hence
\lalign{comm4}{
\gt^{kl}\Db_k\Db_l u^{ij} & = \gt^{00}\delb_0 u_0^{ij} + \gt^{0I}\delb_0 u_I^{ij} + \gt^{0I} \delb_I u_0^{ij}
+ \gt^{IJ}\delb_I u_J^{ij} + \gt^{kl}(-\gammab_{kl}^m u_m^{ij} + 2\gammab_{km}^{(i}u_l^{j)m})\notag \\
& = \gt^{00}\delb_0 u_0^{ij} + 2\ep u^{0I}\delb_I u_0^{ij} + \gt^{IJ} \delb_I u_J^{ij} + \ep d_1^{ij} + \ep^2 d_2^{ij},
\label{comm4.1}
}
where
\lalign{ddef}{
d_1^{ij} &= \hb^{kl}\left(-\frac{1}{\ep}\gammab_{kl}^m u_m^{ij} + \frac{2}{\ep}\gammab_{km}^{(i}u_l^{j)m}\right) \label{ddef.1}
\intertext{and}
d_2^{ij} &= u^{kl}\left(-\frac{1}{\ep}\gammab_{kl}^m u_m^{ij} + \frac{2}{\ep}\gammab_{km}^{(i}u_l^{j)m}\right)+ u^{0I}\left(\frac{2}{\ep}\gammab_{Im}^{(i}u_0^{j)m} -  \frac{2}{\ep}\gammab_{0m}^{(i}u_I^{j)m}
- \frac{2}{\ep^2}\Rcb_{0Im}{}^{(i}u^{j)m}\right).
}
Setting
\leqn{edef}{
e^{ij}_J = \frac{2}{\ep}\gammab_{Jm}^{(i}u_0^{j)m} -  \frac{2}{\ep}
\gammab_{0m}^{(i}u_J^{j)m}
- \frac{2}{\ep^2}\Rcb_{0Jm}{}^{(i}u^{j)m},
}
equations \eqref{comm3} and \eqref{comm4.1} can be used to write the reduced Einstein equations \eqref{redeqns} in the
following first order form
\lalign{fo1}{
-a(t)\gt^{00}\del_t u_0^{ij} &= 2a(t)u^{0I}\del_I u_0^{ij} + \frac{1}{\ep}a(t)\gt^{IJ}\del_I u_J^{ij}
 + a(t)\bigl(c_1^{ij}  + d_1^{ij}\bigr)  \notag \\
& \qquad + \ep a(t)\bigl(a_1^{ij}+ a_2^{ij}+ a_3^{ij}+c_2^{ij}+d_2^{ij}\bigr) +  4\ep a(t)\left(\Tcb^{ij}-\frac{|\gb|}{|\hb|}\Tb^{ij}\right), \label{fo1.1} \\
a(t)\gt^{IJ}\del_t u_J^{ij} &= \frac{1}{\ep}a(t)\gt^{IJ}\del_J u_0^{ij} + a(t)\gt^{IJ}e^{ij}_J, \label{fo1.2} \\
\intertext{and}
\del_t u^{ij} &= u_0^{ij} - \frac{2}{\ep}\gammab^{(i}_{0k}u^{j)k}. \label{fo1.3}
}

For their definition, the reduced Einstein equations \eqref{fo1.1}-\eqref{fo1.3} require that the matrix $\gt^{ij}$ is invertible. For fixed
 $-\sqrt{2/(3\mu_0)} < \tau_0 < 0$ and $\tau_1>0$, it
is clear from \eqref{flrw3} that
\eqn{abounds}{
0<a(\tau_0)\leq a(t) \leq a(T_0) \quad \forall \; t\in [\tau_0,\tau_1].
}
This implies that the set
\eqn{Vcaldef}{
\Vc_{\tau_0,\tau_1} = \{ \, (r^{ij}) \in \Mbb{4} \, |\, \det(\hb^{ij}+r^{ij})>0 \quad \forall\; t\in[\tau_0,\tau_1]\, \}
}
is open and  contains the origin $(r^{ij})=0$, and moreover, that the reduced Einstein equations \eqref{fo1.1}-\eqref{fo1.3} are well defined
for all $t\in (\tau_0,\tau_1)$ and $(\ep u^{ij})\in \Vc_{\tau_0,\tau_1}$.

\subsect{eul}{Regularized Euler equations}

In the coordinates $(\xb^i)$, the Euler equations are given by
\leqn{eul1}{ \nablab_i \Tb^{ij} =0} where $\Tb^{ij} = (\rho + \ep^2
p) \vb^i\vb^j + p \gb^{ij}$ and the fluid velocity $\vb^i$ is
normalized according to
\leqn{eul2}{\vb_i\vb^i =
-\frac{1}{\ep^2}\, .}
To derive a symmetric hyperbolic system for the Euler system, we follow the method of
\cite{BK07} and differentiate \eqref{eul2} to get
\leqn{eul3}{\vb_i \nablab_j \vb^i = 0}
which in turn implies
\leqn{eul4}{\vb^{j}\vb_i \nablab_j \vb^i = 0.}
Writing out
\eqref{eul1} explicitly, we have
\leqn{eul5}{(\delb_i\rho
+\ep^2\delb_i p)\vb^i\vb^j + (\rho+\ep^2 p)(\vb^j\nablab_i\vb^i
+\vb^i\nablab_i\vb^j) + \gb^{ij}\delb_i p = 0\, . }
Next, we observe that the operator
\eqn{proj}{L^j_i = \delta^j_i + \ep^2 \vb^j\vb_i}
projects into
subspace orthogonal to the fluid velocity $\vb^i$, i.e.
$L^j_{i}L^i_k = L^j_k $ and $L^j_i\vb^i=0$.
Applying this operator to project \eqref{eul5} into components parallel
and orthogonal to $\vb^i$ yields, after using the relations
\eqref{eul2}-\eqref{eul4}, the following system\footnote{Recall that we are assuming that the fluid satisfies an
adiabatic equation of state $p=f(\rho)$ (see \eqref{eos}).}
\lgath{eul6}{
\frac{f'(\rho)}{(\rho+\ep^2 f(\rho))^2}\vb^i\delb_i
\rho + \frac{f'(\rho)}{\rho+\ep^2 f(\rho)} L^i_j\nablab_i\vb^j = 0 \, ,\label{eul6.1} \\
M_{ij}\vb^k\nablab_k \vb^j + \frac{f'(\rho)}{\rho+\ep^2 p} L^i_j \delb_i
\rho = 0\, ,\label{eul6.2}} where \eqn{Mdef}{ M_{ij} = \gb_{ij} +
2\ep^2 \vb_i\vb_j . }

As discussed in the introduction, we need to introduce a new fluid four-vector
by
\leqn{wdef}{
w^i = \vb^i-\xib^i = \vb^i-\frac{\delta^i_0}{\ep}.
}
So, letting
\leqn{wvdef}{
\wv = (\rho,w^i)^T
}
allows us to write the system \eqref{eul6.1}-\eqref{eul6.2} as
\leqn{eul7}{
A_M^0\del_t \wv = A_M^I\del_I \wv + F_M,
}
where
\lalign{eul8}{
A_M^0 & = \begin{pmatrix} \frac{f'(\rho)(1+\ep w^0)}{(\rho+\ep^2 f(\rho))^2} &  \frac{ \ep f'(\rho)}{\rho+\ep^2 f(\rho)} L^0_j \\
\frac{\ep f'(\rho)}{\rho+\ep^2 f(\rho)} L^0_i &  M_{ij}(1+\ep w^0) \end{pmatrix}, \label{eul8.1} \\
A_M^I & = \begin{pmatrix} -\frac{f'(\rho)w^I}{(\rho+\ep^2 f(\rho))^2} &  -\frac{ f'(\rho)}{\rho+\ep^2 f(\rho)} L^I_j \\
-\frac{f'(\rho)}{\rho+\ep^2 f(\rho)} L^I_i &  -M_{ij}w^I \end{pmatrix}, \label{eul8.2}
\intertext{and}
F_M & = \begin{pmatrix} \frac{f'(\rho)}{\rho+\ep^2 f(\rho)} L^i_j\bigl(\gammab^{j}_{il}-\Gammab^{j}_{il}\bigr)\vb^l
-\frac{f'(\rho)}{\rho+\ep^2 f(\rho)}L^i_j\gammab^{j}_{il}\vb^l \\
M_{ij}\bigl(\gammab^{j}_{kl}-\Gammab^{j}_{kl}\bigr)\vb^k\vb^l - M_{ij} \gammab^{j}_{kl} \vb^k\vb^l
\end{pmatrix}. \label{eul8.3}
}
Next, a straightforward calculation using \eqref{utog} and \eqref{christ} shows that
\alin{eul9}{
& M_{ij}  = \hb_{ij} + 2\delta_i^0\delta_j^0 + \ep m_{ij}(\ep,t,u^{kl},w^k), \\
&L^{i}_{j}  = \delta^{i}_j - \delta^i_0\delta^0_j + \ep\ell_i^j(\ep,t,u^{kl},w^k), \\
& L^i_j\bigl(\gammab^{j}_{il}-\Gammab^{j}_{il}\bigr)\vb^l = \ep q_0(\ep,t,u^{ij},u^{ij}_k,w^i), \\
& L^i_j\gammab^{j}_{il}\vb^l = \frac{3}{2}\frac{a'(t)}{a(t)}
+ \ep q_1(\ep,t,u^{ij},w^i),\\
& \bigl(\gammab^{j}_{kl}-\Gammab^{j}_{kl}\bigr)\vb^k\vb^l = -u^{j0}_0 + \frac{1}{4}\delta^{j0}(3 u^{00}_0 - a(t)\delta_{KL}u^{KL}_0) \notag \\
&\text{\hspace{2.0cm}}
-\frac{1}{4}\delta^{jI}\left(\frac{1}{a(t)} u_I^{00} + \delta_{KL} u^{KL}_I\right) + \ep q^j_0(\ep,t,u^{ij},u^{ij}_k,w^i),\\
& \gammab^{j}_{kl} \vb^k\vb^l = \frac{a'(t)}{a(t)}\delta^j_I w^I + \ep \left[ \frac{1}{\ep} \gammab^j_{kl}w^k w^l\right],
}
where the maps $m_{ij}$, $\ell^j_i$, $q_0$, $q_1$, and $q_0^j$ are analytic in all their variables
provided that $t\in (\tau_0,\tau_1)$ and $(\ep u^{ij})\in \Vc_{\tau,\tau_1}$.
Using these expressions, we can decompose $A^0_M$, $A^I_M$, and $F_M$ as
\lalign{AFdec1}{
A^0_M &= A^0_{M,0} + \ep A^0_{M,1}(t,\ep,u^{ij},\rho,w^i), \label{AFdec1.1} \\
A^I_M &= A^I_{M,0}+\ep A^I_{M,1}(t,\ep,u^{ij},\rho,w^i), \label{AFdec1.2} \\
F_M & = F_{M,0} + \ep F_{M,1}(t,\ep,u^{ij},u^{ij}_k,\rho,w^i), \label{AFdec1.3}
}
where
\lalign{AFdec2}{
A^0_{M,0} & = \begin{pmatrix} \frac{f'(\rho)}{\rho^2} & 0 \\ 0 & \hb_{ij}+ 2\delta_i^0\delta_j^0\end{pmatrix}, \label{AFdec2.1} \\
A^I_{M,0} & = \begin{pmatrix} -\frac{f'(\rho)}{\rho^2}w^I & -\frac{f'(\rho)}{\rho} \delta^I_j \\
                              -\frac{f'(\rho)}{\rho}\delta^I_i & -( \hb_{ij}+ 2 \delta^0_i\delta^0_j)w^I \end{pmatrix},
                              \label{AFdec2.2} \\
F_{M,0} & = \begin{pmatrix} -\frac{f'(\rho)}{\rho}\frac{3}{2}\frac{a'(t)}{a(t)} \\ ( \hb_{ij}+ 2\delta^0_i\delta^0_j)\left[-u^{j0}_0 + \frac{1}{4}\delta^{j0}(3 u^{00}_0 -
a(t)\delta_{KL}u^{KL}_0)
-\frac{1}{4}\delta^{jI}\left(\frac{1}{a(t)} u_I^{00} + \delta_{KL} u^{KL}_I\right) - \frac{a'(t)}{a(t)}\delta^j_I w^I \right]\end{pmatrix}, \label{AFdec2.3}
}
and the maps $A^0_{M,1}$, $A^I_{M,0}$, $F_{M,1}$ are analytic in all their variables
provided that  $t\in (\tau_0,\tau_1)$ and $(\ep u^{ij})\in \Vc_{\tau_0,\tau_1}$.

\subsect{nloc}{A nonlocal symmetric hyperbolic formulation}

To bring the reduced Einstein equations \eqref{fo1.1}-\eqref{fo1.3} into a form that is suitable to analyze the
limit $\ep\searrow 0$, we replace the $u^{ij}_J$ with the variables
\leqn{Wdef1}{
W^{ij}_I = u^{ij}_I - \del_I\Phi^{ij},
}
where the $\Phi^{ij}$ satisfy
\leqn{Phidef}{
(\delta^{IJ}+ a(t)\ep u^{IJ})\del_I\del_J \Phi^{ij} =  4\ep^2 a(t)\Pi\left(\frac{|\gb|}{|\hb|}\Tb^{ij}-\Tcb^{ij}\right).
}
A short calculation shows that
\leqn{sdiff1}{
4\ep^2\left(\frac{|\gb|}{|\hb|}\Tb^{ij}-\Tcb^{ij}\right) = 4(\rho-\mu(t))\delta^{i}_0\delta^{j}_0
+ \ep \bigl[ 8\delta^{(i}_0 \rho w^{j)}+ 4\delta^i_0\delta^j_0 \rho \hb_{kl}u^{kl} \bigr] + \ep^2 S_0^{ij}(\ep,t,u^{kl},\rho,w^k),
}
where again the map $S_0^{ij}$ is
analytic in all variables provided that
$t\in (\tau_0,T_0)$ and
$(\ep u^{ij}) \in \Vc_{\tau_0,\tau_1}$.
In addition to $\Phi^{ij}$, we will also need the time derivative
\eqn{Phitdef}{
\Phid^{ij} = \del_t\Phi^{ij}
}
which satisfies
\leqn{Phidef1}{
(\delta^{IJ}+ a(t)\ep u^{IJ})\del_I\del_J \Phid^{ij} = -\ep u_4^{IJ}\del_I\del_J\Phi^{ij}
+
4 \ep^2 a(t)\Pi \del_t \left(\frac{|\gb|}{|\hb|}\Tb^{ij}-\Tcb^{ij}\right) +
4 \ep^2 a'(t)\Pi\left(\frac{|\gb|}{|\hb|}\Tb^{ij}-\Tcb^{ij}\right).
}
Using \eqref{eul6.1}-\eqref{eul6.2} to replace the time derivatives of $\rho$ and $w^i$ in favor of
spatial derivatives, we find that
\lalign{sdiff2}{
4\ep^2 \del_t \left(\frac{|\gb|}{|\hb|} \Tb^{ij}-\Tcb^{ij}\right)
 = &-4\left(\del_I(\rho w^I)+\frac{3}{2}\frac{a'(t)}{a(t)}(\rho-\mu(t))\right)
\delta^i_0 \delta^j_0 + \ep \left[ - 8 \del_I(w^I\delta^{(i}_0 \rho w^{j)}) \right. \notag \\
& - \frac{8}{a(t)}\delta^{(i}_0\delta^{j)I}
\del_I f(\rho)
 - 8 \frac{a'(t)}{a(t)}\left(\frac{3}{2}\delta^{(i}_0\rho w^{j)} + \delta^{(i}_0\delta^{j)}_I\rho w^I\right)
+ 8\rho \delta^{(i}_0\beta^{j)} \notag \\
&\left. + 4\delta^i_0\delta^j_0\left( \left(-\del_I(\rho w^I)-\frac{3}{2}\frac{a'(t)}{a(t)}
\rho\right)\hb_{kl} u^{kl} + \rho\left(a'(t)\delta_{IJ} u^{IJ} + \hb_{kl} u_0^{kl}\right) \right) \right] \notag \\
&+ \ep^2 \Bigl[ S_1^{ij}\bigl(\ep,t,u^{kl},\rho,w^k\bigr) + S_2^{ij}\bigl(\ep,t,u^{kl},\rho,w^k,
\del_K \rho,\del_K w^k, u^{kl}_m\bigr)\Bigr] \label{sdiff2.1}
}
where
\eqn{betadef}{
\beta^j = -u^{j0}_0 + \frac{1}{4}\delta^{j0}(3 u^{00}_0 -
a(t)\delta_{KL}u^{KL}_0)
-\frac{1}{4}\delta^{jI}\left(\frac{1}{a(t)} u_I^{00} + \delta_{KL} u^{KL}_I\right),
}
and the maps $S_\alpha^{ij}$ $(\alpha=1,2)$ are analytic in all variables provided that
$t\in (\tau_0,T_0)$,
$(\ep u^{ij}) \in \Vc_{\tau_0,T_0}$, and $1+\ep w^0 > 0$, and $S_2^{ij}$ is linear
in $(\del_K\rho,\del_K w^k, u^{kl}_m)$.

Substituting \eqref{Wdef1} into \eqref{fo1.1}-\eqref{fo1.2} gives
\lalign{fo2}{
-a(t)\gt^{00}\del_t u_0^{ij} &= 2a(t)u^{0I}\del_I u_0^{ij} + \frac{1}{\ep}a(t)\gt^{IJ}\del_I W_J^{ij}
 + a(t)\bigl(c_1^{ij}  + d_1^{ij}\bigr) \notag \\
 &\text{\hspace{3.5cm}}  + \ep a(t)\bigl(a_1^{ij}+ a_2^{ij}+ a_3^{ij}+c_2^{ij}+d_2^{ij}\bigr) +  \phi^{ij}, \label{fo2.1} \\
a(t)\gt^{IJ}\del_t W_J^{ij} &= \frac{1}{\ep}a(t)\gt^{IJ}\del_J u_0^{ij} + a(t)\gt^{IJ}\left[-\del_J \Phid^{ij} + e^{ij}_J\right], \label{fo2.2}
}
where
\leqn{phidef}{
\phi^{ij}(t) = 4\ep a(t) \left\langle 1 \left| \Tcb^{ij}-\frac{|\gb|}{|\hb|}\Tb^{ij} \right.\right\rangle_{L^2}.
}

Differentiating \eqref{phidef} with repsect to $t$ while using \eqref{sdiff1} and \eqref{sdiff2.1}, we find that
\leqn{phieqn}{
\phi^{ij}{}^\prime (t) =  -\frac{1}{2}\phi^{ij} (t) + F^{ij}_\phi - \ep\biggl\langle 1 \biggl|
a'(t)\frac{3}{2}S_0^{ij} + a(t)\bigl(S_1^{ij}+S_2^{ij}\bigr) \biggr\rangle_{L^2},
}
where
\lalign{Fphidef}{
F^{ij}_\phi = & 8 a(t)\ip{1}{\rho w^I}_{L^2}\delta_0^{(i}\delta^{j)}_I - 8a(t)\biggl\langle 1 \biggl| \rho \left(
-\delta^{(i}_0 u^{j)0} + \frac{1}{4}\delta^{(i}_0\delta^{j)0}(3 u^{00}_0 -
a(t)\delta_{KL}u^{KL}_0)\right) \biggr\rangle_{L^2} \notag \\
& + 2\ip{\Pi\rho}{\delta^{(i}_0\delta^{j)I}\bigl(u_I^{00} + a(t) \delta_{KL} u^{KL}_I\bigr) }_{L^2}
- 4 a(t)\ip{1}{\rho\bigl(a'(t)\delta_{IJ}u^{IJ}+ \hb_{kl} u^{kl}_0\bigr)}_{L^2}\delta^i_0\delta^j_0 \label{Fphidef.1}.
}

Next, we define
\leqn{Wdef}{
W = (u_0^{ij},W_I^{ij},u^{ij},\phi^{ij},\wv)^T.
}
Then it follows from \eqref{fo1.3},
\eqref{eul7}, \eqref{AFdec1.1}-\eqref{AFdec1.3}, \eqref{fo2.1}-\eqref{fo2.2}, and \eqref{phieqn} that $W$ satisfies
\leqn{nonloc1}{
A^0 \del_t W = \frac{1}{\ep}C^{I}\del_I W + (A_0^I + \ep A_1^I)\del_I W + F_0 + \ep F_1,
}
where
\lalign{Adef}{
A^0 & = \begin{pmatrix} A^0_G & 0 \\ 0 & A^0_{M,0}+\ep A^0_{M,1} \end{pmatrix} \label{Adef.1} , \\
A^0_G & = \begin{pmatrix} a(t)(1-\ep u^{00}) & 0 & 0 & 0\\
0  & (\delta^{IJ}+\ep a(t) u^{IJ})  & 0  & 0\\
0 & 0 & 1  & 0 \\
0 & 0 & 0 & 1 \end{pmatrix} \label{Adef.2} ,
}
\lalign{Adef1}{
A^I_0 & = \begin{pmatrix} A^I_G & 0 \\ 0 & A^I_{M,0} \end{pmatrix} \label{Adef.3} , \\
A^I_G & = \begin{pmatrix} 2a(t) u^{0I} & a(t) u^{IJ} & 0 & 0 \\
a(t) u^{IJ} & 0 & 0 & 0\\
0 & 0 & 0 & 0 \\
0 & 0 & 0 & 0 \end{pmatrix} \label{Adef.4},  \\
A^I_1 & = \begin{pmatrix} 0 & 0 \\ 0 & A^I_{M,1} \end{pmatrix} \label{Adef.5} ,
}
\lalign{Adef2}{
C^I & = \begin{pmatrix} C^I_G & 0 \\ 0 & 0 \end{pmatrix} \label{Adef.6},\\
C^I_G & = \begin{pmatrix} 0 & \delta^{IJ} & 0 & 0 \\ \delta^{IJ} & 0 & 0 & 0 \\
0 & 0 & 0 & 0 \\
0 & 0 & 0 & 0 \end{pmatrix}, \label{Adef.7}
}
and
\lalign{Fdef}{
F_0 & = \begin{pmatrix} a(t)\bigl(c_1^{ij}  + d_1^{ij}\bigr) + \phi^{ij} \\
                        -\delta^{IJ}\del_J \Phid_0 \delta^i_0\delta^j_0 + \delta^{IJ}e^{ij}_{J}\\
                        u_0^{ij} - \frac{2}{\ep}\gammab^{(i}_{0k}u^{j)k}  \\
                        -\frac{1}{2}\phi^{ij} (t) + F_\phi^{ij}           \\
                             \tilde{F}_{M,0}
                             \end{pmatrix}, \label{Fdef.1} \\
\tilde{F}_{M,0} & = \begin{pmatrix} -\frac{f'(\rho)}{\rho}\frac{3}{2}\frac{a'(t)}{a(t)} \\ ( \hb_{ij}+ 2\delta^0_i\delta^0_j)\tilde{\beta}^j \end{pmatrix}, \label{Fdef.2}
}
\lalign{Fdef1}{
\tilde{\beta}^j & = -u^{j0}_0 + \frac{1}{4}\delta^{j0}(3 u^{00}_0 -
a(t)\delta_{KL}u^{KL}_0)
-\frac{1}{4}\delta^{jI}\left(\frac{1}{a(t)}u_I + \delta_{KL} u^{KL}_I\right) - \frac{a'(t)}{a(t)}\delta^j_I w^I, \label{Fdef.3} \\
\Phi_0 & =  4 a(t) \Delta^{-1} \Pi\bigl(\rho - \mu(t)\bigr) , \label{Fdef.4} \\
\Phid_0 & = -4 a(t)\Pi\left(\del_I(\rho w^I) + \frac{1}{2}\frac{a'(t)}{a(t)} (\rho-\mu(t))\right),
\label{Fdef.5}
}
\lalign{Fdef2}{
F_1 & = \begin{pmatrix}  a(t)\bigl(a_1^{ij}+ a_2^{ij}+ a_3^{ij}+c_2^{ij}+d_2^{ij}\bigr)  \\
-\frac{1}{\ep}\delta^{IJ}\del_J(\Phid^{ij}-\Phid_0\delta^i_0\delta^j_0)  + a(t)u^{IJ}\bigl[ -\del_J\Phid^{ij}+ e_J^{ij}\bigr] \\
          - \biggl\langle 1 \biggl|
a'(t)\frac{3}{2}S_0^{ij} + a(t)\bigl(S_1^{ij}+S_2^{ij}\bigr) \biggr\rangle_{L^2}   \\
          \frac{1}{\ep}\bigl(F_{M,0}-\tilde{F}_{M,0}\bigr)+ F_{M,1} \end{pmatrix}. \label{Fdef.6}
}

\subsect{welldef}{Well-posedness of the nonlocal system}

With the evolution equations in a suitable form, we now verify that the system is well-posed.
To do this, we will repeatedly use the following
elementary facts concerning analytic maps:
\begin{lem} \label{facts} \mnote{[facts]}
Let $X$, $Y$, and $Z$ be Banach spaces with $U\subset X$ and $V\subset Y$ open.
\begin{itemize}
\item[(i)] If $L : X \longrightarrow Y$ is a continuous linear map, then $L \in C^\omega(X,Y)$.
\item[(ii)] If $B : X\times Y \longrightarrow Z$ is a continuous bilinear map, then $B \in C^\omega(X\times Y,Z)$.
\item[(iii)] If $f\in C^\omega(U,Y)$, $g\in C^\omega(V,Z)$ and $\text{\rm ran}(f)\subset V$, then
$g\circ f \in C^\omega(U,Z)$.
\end{itemize}
\end{lem}
We also recall the well-known Multiplication Lemma.
\begin{lem} \label{mlem} \mnote{[mlem]}
Suppose $s_1, s_2 \geq s_3\geq 0$ and $s_3 < s_1+s_2-3/2$. Then there exists a constant $C>0$
such that
\eqn{mlem1}{
\norm{\psi_1 \psi_2}_{H^{s_3}} \leq C \norm{\psi_1}_{H^{s_1}} \norm{\psi_2}_{H^{s_2}}
}
for all $\psi_1\in H^{s_1}$ and $\psi_2\in H^{s_2}$.
\end{lem}
This lemma shows that the bilinear map
\leqn{bmap}{
H^{s_1}\times H^{s_2} \ni (\psi_1,\psi_2) \longmapsto \psi_1 \psi_2 \in H^{s_3}
}
is continuous, and hence analytic, provided that $s_1, s_2 \geq s_3 \geq 0$ and $s_3 < s_1+s_2-3/2$. In particular,
$H^s$ is a Banach algebra for $s>3/2$ which implies that there exists a constant $C_s>0$  such that
\leqn{BA}{
\norm{\psi_1 \psi_2}_s \leq C_s \norm{\psi_1}_{H^s}\norm{\psi_2}_{H^s}
}
for all $\psi_1, \psi_2\in H^{s}$. This can be used to prove the following important proposition concerning analytic maps.
For a proof, see Proposition 3.6 of \cite{Heil}.
\begin{prop} \label{analprop} \mnote{[analprop]}
Suppose $s>3/2$, $F\in C^\omega(B_R(\Rbb^N),\Rbb)$, $C_s$ is the constant from the inequality \eqref{BA}, and that
\eqn{analprop1}{
F(y_1,\ldots,y_N) = F_0 + \sum_{|\alpha|\geq 1} c_\alpha \, y_1^{\alpha_1}y_2^{\alpha_2}\cdots y_N^{\alpha_n}
}
is the powerseries expansion for $F(y)$ about $0$.
Then the map
\eqn{analprop1}{
\bigl(B_R(H^s)\bigr)^N \ni (\psi_1,\psi_2,\ldots,\psi_N) \longmapsto F(\psi_1,\psi_2,\ldots,\psi_N) \in H^s
}
is in $C^\omega\bigl( \bigl(B_{R/C_s}(H^k)\bigr)^N, H^s\Bigr)$, and
\eqn{analprop2}{
F(\psi_1,\ldots,\psi_N) = F_0 + \sum_{|\alpha|\geq 1} c_\alpha \, \psi_1^{\alpha_1}f_2^{\alpha_2}\cdots \psi_N^{\alpha_N}
}
for all $(\psi_1,\ldots,\psi_N)\in \bigl(B_{R/C_s} (H^s)\bigr)^N$.
\end{prop}

The first step in establishing well-posedness is to show that the maps $\Phi$ and $\Phid$ are well-defined and analytic.
\begin{lem} \label{philem} \mnote{[philem]}
Suppose $R>0$ and $s > 3/2$. Then there exists an $\ep_0 > 0$ and
an analytic map
\eqn{philem1}{
(-\ep_0,\ep_0) \times (\tau_0,T_0)\times H^{s} \times H^{s}(\Rbb^4) \times B_R\bigl(H^{s} (\Sbb{4})\bigr)
\ni (\ep,t,\rho,w^i,u^{ij})  \longmapsto (\Phi^{ij}) \in \Ho^{s+2}(\Sbb{4})
}
that satisfies \eqref{Phidef} and
\eqn{philem2}{
\Phi^{ij} \bigl|_{\ep=0} = 4a(t)\delta^{i}_0\delta^{j}_{0}\Delta^{-1}\Pi\bigl(\rho-\mu(t)\bigr).
}
\end{lem}
\begin{proof} First we observe that for a fixed $R>0$,  the Born series
\eqn{Born}{
\Bigl[(\delta^{IJ}+\ep a(t) u^{IJ})\del_{I}\del_J\Bigr]^{-1}=
\Bigl[\id + \ep a(t)\Delta^{-1}u^{IJ}\del_I\del_J\Bigr]^{-1}\Delta^{-1} = \sum_{n=0}^\infty \ep^n (-1)^n a(t)^n
(\Delta^{-1}u^{IJ}\del_I\del_J\bigl)^n \Delta^{-1},
} the Multiplication Lemma \ref{mlem}, and the invertibility
 of the Laplacian $\Delta : \Ho^{s+2}\rightarrow \Ho^{s}$ show that there exists an $\ep_0>0$ such that the map
\leqn{philem3}{
(-\ep_0,\ep_0)\times (\tau_0,T_0)\times B_R(H^{s+2}(\Sbb{3}))\times \Ho^{s}(\Sbb{4}) \ni (\ep,t,u^{IJ},\Psi^{ij}) \longmapsto
\Bigl[\bigl(\delta^{IJ} +\ep a(t) u^{IJ}\bigr)\del_I\del_J\Bigr]^{-1}\Psi^{ij} \in \Ho^{s+2}(\Sbb{4})
}
is well defined and analytic. Also, by Lemma \ref{facts} and Proposition \ref{analprop}, we see that (shrinking $\ep_0$ if neccessary)
the map
\leqn{philem4}{
(-\ep_0,\ep_0) \times (\tau_0,T_0)\times H^{s} \times H^{s}(\Rbb^4) \times H_R^{s} (\Sbb{4})
\ni (\ep,t,\rho,w^i,u^{ij} ) \longmapsto 4\ep^2 a(t)\Pi\left(\frac{|\gb|}{|\hb|}\Tb^{ij}-\Tcb^{ij}\right)  \in H^{s}(\Sbb{4})
}
is well-defined and analytic. The proof then follows directly from composing the two analytic maps \eqref{philem3}-\eqref{philem4}, i.e.
\eqn{philem5}{
\Phi^{ij} = \Bigl[\bigl(\delta^{IJ} +\ep a(t) u^{IJ}\bigr)\del_I\del_J\Bigr]^{-1}\left[4\ep^2 a(t)\Pi\left(\frac{|\gb|}{|\hb|}\Tb^{ij}-\Tcb^{ij}\right)\right],
}
which is again analytic by Lemma \ref{facts}.
\end{proof}

\begin{lem} \label{phidlem} \mnote{[phidlem]}
Suppose $R>0$ and $s > 3/2$. Then there exists an $\ep_0 > 0$ and
an analytic map
\eqn{phidlem1}{
(-\ep_0,\ep_0) \times (\tau_0,T_0)\times H^{s} \times H^{s}(\Rbb^4) \times B_R\bigl(H^{s} (\Sbb{4})\bigr) \times H^{s}(\Sbb{4})
\ni (\ep,t,\rho,w^i,u^{ij},u^{ij}_k)  \longmapsto (\Phid^{ij}) \in \Ho^{s+1}(\Sbb{4})
}
that satisfies \eqref{Phidef1} and
\eqn{phidlem2}{
\Phid^{ij} \bigl|_{\ep=0} = -4a(t)\delta^{i}_0\delta^{j}_{0}\Delta^{-1}\Pi\left(\del_I(\rho w^I)+\frac{1}{2}\frac{a'(t)}{a(t)}
(\rho-\mu(t))\right).
}
\end{lem}
\begin{proof}
The proof follows from  a routine adaptation of the proof of Lemma \ref{philem}.
\end{proof}

Next, we introduce the space
\eqn{Hcal}{
\Hc_R^{s} = H^{s}(\Sbb{4})\times \bigl(H^{s}(\Sbb{4})\bigr) \times B_R\bigl(H^{s}(\Sbb{4})\bigr) \times \Sbb{4} \times
H^{s} \times B_R(H^{s}) \times H^{s}(\Rbb^3)
}
and let $\Hc^s = \Hc_\infty^s$.

\begin{lem} \label{Flem} \mnote{[Flem]}
Suppose $R>0$ and $s>3/2$. Then there exists an $\ep_0>0$ such that the maps
\eqn{Flem1}{
(\tau_0,T_0)\times (-\ep_0,\ep_0)\times \Hc_R^{s} \ni  (t,\ep,u_0^{ij},W_I^{ij},u^{ij},\phi^{ij},\rho,w^0,w^I)^T \longmapsto F_\alpha \in
\Hc^{s+1} \quad (\alpha=0,1)
}
are analytic.
\end{lem}
\begin{proof}
Follows directly from Proposition \ref{analprop} and Lemmas \ref{facts}, \ref{philem}, \ref{phidlem}.
\end{proof}

With the analyticity of the maps $F_0, F_1$ established, local existence, uniqueness, and continuation of solutions to the nonlocal symmetric hyperbolic system \eqref{Wdef} follow from standard arguments (see for example \cite{TayIII}, Chapter 16). In particular, we can apply the local existence results of Schochet \cite{Scho86,Scho88} (see also \cite{KM82,Kreiss80}) to obtain the existence of solution to \eqref{Wdef}
on spacetime regions of the form $D=[0,T)\times \Tbb^3$ where $T$ is independent of $\ep$. This will be discussed in detail in Section \ref{proof}. 
\sect{init}{Initialization}

\subsect{const}{Constraint equations}

In order to solve the initial value problem for the Einstein equations, we must first construct initial data that satisfies the
following constraint equations on the initial hypersurface defined by $t=0$:
\lalign{ceqns}{
\bigl(\Gb^{0j}-2\ep^4 \Tb^{0j}\bigr)\bigl|_{t=0} &= 0 && \text{(gravitational constraints)}, \label{ceqns.1} \\
 \Db_i \ub^{ij}\bigl|_{t=0} &= 0 && \text{(harmonic gauge condition)} \label{ceqns.2},
\intertext{and}
\left(\gb_{ij}\vb^{i} \vb^{j} + \frac{1}{\ep^2}\right)\Bigl|_{t=0} &= 0 && \text{(fluid 4-velocity normalization)}.
\label{ceqns.3}
}
A short calculation using \eqref{hcurv.1}-\eqref{hcurv.2} shows that the harmonic condition \eqref{ceqns.2} is equivalent to
\lalign{ceqns1}{
\del_t\ub^{00} & = -\frac{1}{\ep} \del_I \ub^{I0} -\frac{a'(t)}{2a(t)}\bigl(3\ub^{00} + a(t)\delta_{IJ}\ub^{IJ}\bigr), \label{ceqns1.1}\\
\del_t\ub^{0J} & = -\frac{1}{\ep} \del_I \ub^{IJ} - \frac{5 a'(t)}{2 a(t)} \ub^{0J}. \label{ceqns1.2}
}

Using formulas \eqref{utog} and \eqref{Gb1}-\eqref{Gb2.7}, it is not difficult to verify that the
gravitational constraint equations \eqref{ceqns.1} do not involve second order time
derivatives. In fact, using \eqref{ceqns1.1}-\eqref{ceqns1.2}, the top derivative terms can be expanded as
\lalign{symbol1}{
\gh^{kl}\Db_k\Db_l & \ub^{0j} + b^{0j} = \frac{1}{a(t)}\Delta \ub^{0j}
 -\delta^j_0\del_I\del_J\ub^{IJ} + \ep \left[\delta^{j}_J \left(
\del_I \del_t \ub^{IJ}
+ \frac{a'(t)}{a(t)^2}\delta^{JK}\del_K\ub^{00} \right. \right. \notag \\
&\left. \left. +\frac{a'(t)}{a(t)} \del_{I}\ub^{IJ}\right)
   +\delta^j_0\frac{a'(t)}{2 a(t)}\del_I\ub^{I0} \right] + \ep^2
  \Bigl[ f^j_{0}\bigl(\ep,t,\ub^{ij}\bigr)+f_1\bigl(\ep,t,\ub^{ij},\del_K\del_L\ub^{ij},\del_K \ub^{ij},\del_t\ub^{ij}\bigr)
  \Bigr] \label{symbol1.1}
}
where $f_0$ and $f_1$ are  analytic in all of their variables, and $f_2$ is linear
in $(\del_K\del_L\ub^{ij},\del_K \ub^{ij},\del_t\ub^{ij})$.
Together, \eqref{utog}, \eqref{Gb1}-\eqref{Gb2.7}, \eqref{sdiff1}, and \eqref{symbol1.1} show that
the constraint equations (when evaluated at $t=0$) can be written as
\lalign{ceqn2}{
\Delta \ub^{0j}& -\delta^j_0 a_0\Bigl[4(\rho-\mu)+ \del_I\del_J\ub^{IJ}\Bigr] + \ep a_0 \left[ -4\rho(w^0\delta^j_0 + w^j) + \delta^{j}_J \left(
\del_I \del_t \ub^{IJ}
+ \frac{a'(0)}{a_0^2}\delta^{JK}\del_K\ub^{00}  \right. \right. \notag \\
& \left. \left. +\frac{a'(0)}{a_0} \del_{I}\ub^{IJ}\right)
   +\delta^j_0\frac{a'(0)}{2 a_0}\del_I\ub^{I0}
 \right] + \ep^2 \Bigl[f^j_{2}\bigl(\ep,\ub^{ij},\del_K\del_L\ub^{ij},\del_K \ub^{ij},\del_t\ub^{ij}\bigr) \notag \\
  & \qquad + f^j_3\bigl(\ep,\ub^{ij},\del_t\ub^{ij},\del_K \ub^{ij}\bigr)
+ f^j_4\bigl(\ep,\ub^{ij},\rho,w^0,\rho w^I\bigr)\Bigr] = 0, \label{ceqn2.1} }
where
for any $R>0$ there exists an $\ep_0 > 0$ such that the maps $f_\alpha$ $(\alpha=2,3,4)$
are analytic in all their variables provided $|\ep|<\ep_0$, and $|\ub^{ij}|<R$, $ f^j_{2}$ is linear in $(\del_K\del_L\ub^{ij},\del_K\ub^{ij},\del_t\ub^{ij})$, and $f^j_2$  is quadratic in
$(\del_t\ub^{ij},\del_K \ub^{ij})$.
Also, an easy calculation using \eqref{utog} shows that \eqref{ceqns.3} takes the form
\leqn{ceqn3}{
w^0  + \frac{1}{\ep} + \frac{\gb_{0J}w^J+\sqrt{\ep^2(\gb_{0J}w^J)^2-\gb_{00}(\ep^2\gb_{IJ}w^I w^J+1)}}{\ep g_{00}}
= w^0 - \ep f_0\bigl(\ep, u^{ij},w^{I}\bigr) = 0,
}
where the map $f_0$ is
analytic provided $|\ep|<\ep_0$, $|\ub^{ij}|<R$, and $|w^I|<R$.

\subsect{IJredeqns}{$IJ$-components of the reduced Einstein Equations}

The $IJ$-components of the FLRW wave operator acting on $\ub^{ij}$ are given by
\lalign{wave}{
\hb^{kl}\Db^{k}\Db_{l} \ub^{IJ} = -\ep^2\del_t^2 & \ub^{IJ} + \frac{1}{a(t)}\Delta \ub^{IJ}
+\ep \frac{a'(t)}{a(t)^2}\bigl(\del_I \ub^{0J} + \del_J \ub^{0I}\bigr)  \notag \\
& + \ep^2\left[
-\frac{7}{2}\frac{a'(t)}{a(t)} \del_t \ub^{IJ} - \left( \del_t\left( \frac{a'(t)}{a(t)}\right) + 2
\frac{a'(t)^2}{a(t)^2} \right)\ub^{IJ} + \frac{a'(t)^2}{2 a(t)^3} \ub^{00}\right].
\label{wave.1}}
A calculation involving the harmonic conditions \eqref{ceqns1.1}-\eqref{ceqns1.2}, and formulas \eqref{utog},
\eqref{Gb2.1}-\eqref{Gb2.7}, \eqref{redGb}, \eqref{sdiff1}, \eqref{wave.1}
shows that the $IJ$-components of the reduced Einstein equations can be written as
\lalign{IJred}{
\Delta \ub^{IJ}  +  \ep\del_I&\left( \frac{a'(t)}{a(t)} u^{0J}\right) +
  \ep\del_J\left(\frac{a'(t)}{a(t)} u^{0I}\right) - \ep^2 a(t)\left[  \del_t^2\ub^{IJ}
   -\frac{7}{2}\frac{a'(t)}{a(t)} \del_t \ub^{IJ} + \rho w^I w^J + \frac{f(\rho)}{a(t)}\delta^{IJ}  \right] \notag \\
& + \ep^2\Bigl[
p_0^{IJ}(t,\ep,\ub^{ij},\ep^2\del_t^2 \ub^{KL}, \ep\del_t \del_M\ub^{KL},\del_M\del_N\ub^{MN},\del_M\ub^{ij},\ep\del_t\ub^{ij},\ep \ub^{ij})
 \notag \\
& + p_1^{IJ}(t,\ep,\ep^2\ub^{ij},\del_M \ub,\ep\del_t\ub^{ij},\ep\ub^{ij}) + p^{IJ}_2(t,\ep,\ub^{ij}) \Bigr]+
\ep^3 p_3^{IJ}(t,\ep,\ub^{ij},\rho,w^K), \label{IJred.1}
}
where for any $R>0$ there exist an $\ep_0>0$ such that the maps $p_\alpha$ $(\alpha=0,1,2,3)$
are analytic in all variables
provided $-\tau_0< t < \tau_1$, $|\ep|<\ep_0$, and $|\ub^{ij}| < R$. Furthermore,
$p^{IJ}_0$ is linear in $(\ep^2\del_t^2 \ub^{KL}, \ep\del_t \del_M\ub^{KL},\del_M\del_N\ub^{MN},\del_M\ub^{ij},\ep\del_t\ub^{ij},\ep \ub^{ij})$, $p^{IJ}_1$ is quadratic in $(\del_K \ub^{ij},\ep\del_t\ub^{ij},\ep\ub^{ij})$, and $p_2^{IJ}$ is linear
in $(\ub^{ij})$,

\subsect{dusteqns}{The Euler equations}

Directly from equation \eqref{eul7} and formulas \eqref{AFdec1.1}-\eqref{AFdec2.3}, it follows that
\lalign{dust}{
\del_t w^i & = -w^I\del_I w^i -\frac{f'(\rho)}{a(t)\rho}\delta^{iJ}\del_J\rho  - \frac{a'(t)}{a(t)} \delta^{i}_J w^J -\frac{1}{4} \delta^{iJ}\left(\frac{1}{a(t)}\del_J \ub^{00}+ \delta_{KL}\del_J\ub^{KL}\right)
-\notag \\
& \qquad \qquad \ep\Bigl[
q^i_0\bigl(t,\ep,\rho,w^j,\ub^{jk}\bigl)+q^i_1\bigl(t,\ep,\rho,w^j,\ub^{jk},\del_I\rho, \del_I w^j,\del_t\ub^{jk},\del_I u^{jk} \bigr)\Bigr] , \label{dust.1} \\
\del_t \rho &= -\del_I(\rho w^I) - \frac{3}{2}\frac{a'(t)}{a(t)}\rho +  \notag \\
& \qquad \qquad \ep\Bigl[
q_0\bigl(t,\ep,\rho,w^j,\ub^{jk}\bigl)+q_1\bigl(t,\ep,\rho,w^j,\ub^{jk},\del_I\rho, \del_I w^j,\del_t\ub^{jk},\del_I u^{jk} \bigr)\Bigr], \label{dust.2}
}
where for any $R>0$ there exists an $\ep_0>0$ such that the maps $q_\alpha, q^i_\alpha$ $(\alpha=0,1)$
are analytic in all variables
provided $\tau_0 < t < \tau_1$, $|\ep|<\ep_0$, $|\ub^{ij}| < R$, and $|w^0| < R$, and
$q_1, q^{j}_1$ are linear in $(\del_I\rho, \del_I w^j,\del_t\ub^{jk},\del_I u^{jk})$.

\subsect{hotd}{Higher order time derivatives}

As discussed in the introduction, Kreiss's bounded derivative principle requires us to calculate higher order time derivatives
of $\ub^{ij}$, $\rho$, and $w^i$. The fact that the constraint equations must be satisfied
complicates this task, and we find it advantageous to introduce the following rescaled variables:
\lgath{tvars1}{
\delt{\ell} \ub^{IJ} = y_\ell^{IJ} + \ep^2 \ut_\ell^{IJ}, \quad
\delt{\ell} \ub^{0J}  = \ep\bigl(\delta^0_\ell y^{0J}_0  + \ut_\ell^{0J}\bigr), \quad
\delt{\ell} \ub^{00}  = \delta^0_\ell y^{00}_0 + \ut_\ell^{00}, \quad (\ell \geq  0) \label{tvars1.1} \\
\rho|_{t=0} = \mu_0+\lambda_0+ \rhot_0, \quad
w^0|_{t=0} =  \wt_0^0, \quad
(\rho w^I)|_{t=0} =  z^I_0 + \wt_0^I, \label{tvars1.2} \\
\delt{\ell} \rho  = \rhot_\ell, \AND
\delt{\ell} w^i  = \wt_\ell^i \quad (\ell \geq 1), \label{tvars1.3}
}
where $y_\ell^{IJ}$, $z^I_0$, $\lambda_0$
are constants, and
\leqn{tvars3}{
\int_{\Tbb^3} \rhot_0 \, d^3x = \int_{\Tbb^3}\wt_0^I\, d^3x = \int_{\Tbb^3} \ut^{0j}_0\, d^3x = \int_{\Tbb^3} \ut^{IJ}_{\ell}\, d^3 x = 0 \quad (\ell \geq 0).
}
We also define
\lgath{tvars4}{
\utv^{ij}_\ell = (\ut_0^{ij},\ut_1^{ij},\ldots,\ut^{ij}_\ell), \quad \yv^{IJ}_{\ell+2}  = (y^{IJ}_2,\ldots, y^{IJ}_{\ell+2}), \label{tvars4.1} \\
\rhotv^{IJ}_{\ell+1} = (\rhot_1,\rhot_2,\ldots,\rhot_{\ell+1}), \AND \wtv^i_{\ell+1}  = (\wt_1^i,\wt_2^i,\ldots,\wt^i_{\ell+1}). \label{tvars4.4}
}

Differentiating the harmonic conditions \eqref{ceqns1.1}-\eqref{ceqns1.2}, the $IJ$ components of the reduced Einstein equations \eqref{IJred.1},
and the Euler equations \eqref{dust.1}-\eqref{dust.2} with respect to time and evaluating at $t=0$ while using the variables \eqref{tvars1.1}-\eqref{tvars1.3}
yields the following system
of equations \lalign{tvars5}{
&\Delta \ut^{00}_0 -4a_0(\rhot_0 + \lambda_0) - \ep F^0\bigl(\ep,y_0^{ij},y_1^{IJ},\utv_1^{ij},\del_x^{\alpha+1}\ut^{ij},\lambda_0,
\rhot_0,z_0^I,\wt_0^I\bigr)  = 0,\label{tvars5.1} \\
&\Delta \ut^{0J}_0 - 4a_0(z^J_0+\wt^J_0) - \ep F^J\bigl(\ep,y_0^{ij},y_1^{IJ},\utv_1^{ij},\del_x^{\alpha+1}\ut^{ij},\lambda_0,
\rhot_0,z_0^I,\wt_0^I\bigr)  = 0, \label{tvars5.2} \\
&\Delta \ut_\ell^{IJ} - a_0 y^{IJ}_{\ell+2} - P_{0,\ell}^{IJ}\bigl(y_0^{00},\yv^{KL}_{\ell+1},\del_K\utv^{00}_\ell,\utv^{00}_\ell,\lambda_0,\rhot_0,\rhotv_\ell,
z_0^K,\wt^K_0,\wtv^K_\ell\bigr) \notag \\
&\quad -\ep^1 P_{1,\ell}^{IJ}\bigl(\ep,y_0^{ij},y_1^{IJ},\yv_{\ell+2}^{KL},\del_x^\alpha \utv^{ij}_{\ell},\ut^{ij}_{\ell+1},
\del_x \ut^{KL}_{\ell+1},
\del_M\del_N \ut^{KL}_\ell, \ut^{KL}_{\ell+2},\lambda_0,\rhot_0,\rhotv_\ell,z_0^K,\wt_0^K,\wtv^K_\ell \bigr)  =  0, \label{tvars5.3} \\
&\ut^{00}_{\ell+1} + \del_I\ut^{I0}_\ell + \sum_{p=0}^{\ell}\binom{\ell}{p}\left[\frac{3}{2}
\frac{d^{{p+1}}\ln(a)}{dt^{p+1}\;}
\Bigl|_{t=0} \,
\bigl(\ut^{00}_{\ell-p} + \delta_{p0}y_0^{00}\bigr) \right. \notag \\
& \qquad \qquad \left. + \frac{1}{2} \frac{d^{{p+1}}a}{dt^{p+1}\;}
\Bigl|_{t=0}  \,\delta_{IJ}\bigl(y^{IJ}_{\ell-p}+\ep^2 \ut^{IJ}_{\ell-p}\bigr)\right] = 0, \label{tvars5.4} \\
&\ut^{0J}_{\ell+1} + \del_I \ut^{IJ}_\ell + \ep \sum_{p=0}^{\ell}\binom{s}{p}\frac{5}{2}\frac{d^{p+1}\ln(a)}{dt^{p+1}\;}
\Bigl|_{t=0}\, \bigl(\ut^{0J}_{\ell-p}+\delta_{p0} y_0^{0J}\bigr) = 0, \label{tvars5.5}
}
\lalign{tvars5s}{
& \wt_0 -\ep Q^0_0(\ep,y_0^{IJ},\ut^{jk}_0,z_0^K,\wt_0^J) = 0, \label{tvars5.6} \\
& \wt_{\ell+1}^i - Q^i_{0,\ell}\bigl(\lambda_0,\del_x^{\alpha}\rhot_0,\del_x^{\alpha}\rhotv_\ell,z_0^I,
\del_x^\alpha \wt^j,\del_x^\alpha \wtv^j_\ell \bigr) \notag \\
& \qquad \qquad -
\ep Q^i_{1,\ell} \bigl(\ep,y_0^{ij},y_1^{IJ},\yv^{IJ}_{\ell+1},\utv^{ij}_{\ell+1},\del_K\utv^{jk}_\ell,
\lambda_0,\del_x^{\alpha} \rhot_0,\del_x^{\alpha}\rhotv_\ell,z_0^I,
\del_x^\alpha \wt^j,\del_K \utv^{00}_\ell\bigr) = 0,
\label{tavars5.7}
\intertext{and}
&\rhot_{\ell+1}^i - Q_{0,\ell}\bigl(\lambda_0,\del_x^{\alpha}\rhot_0,\del_x^{\alpha}\rhotv_\ell,z_0^I,
\del_x^\alpha \wt^j,\del_x^\alpha \wtv^j_\ell \bigr) \notag \\
& \qquad \qquad -
\ep Q_{1,\ell} \bigl(\ep,y_0^{ij},y_1^{IJ},\yv^{IJ}_{\ell+1},\utv^{ij}_{\ell+1},\del_K\utv^{jk}_\ell,
\lambda_0,\del_x^{\alpha} \rhot_0,\del_x^{\alpha}\rhotv_\ell,z_0^I,
\del_x^\alpha \wt^j\bigr) = 0,  \label{tvars5.8}
}
where $0\leq |\alpha| \leq 1$. Next, we set
\lalign{tvars6}{
&\Psi_0^{00}  = \Delta\ut^{00}_0 -4a_0\rhot_0 - \ep \Pi F^0\bigl(\ep,y_0^{ij},y_1^{IJ},\utv_1^{ij},\del_x^{\alpha+1}\ut^{ij},\lambda_0,
\rhot_0,z_0^I,\wt_0^I\bigr) \label{tvars6.1} \\
&\Psi_0^{0J}  = \Delta \ut^{0J}_0 - 4a_0\wt^J_0 - \ep \Pi F^J\bigl(\ep,y_0^{ij},y_1^{IJ},\utv_1^{ij},\del_x^{\alpha+1}\ut^{ij},\lambda_0,
\rhot_0,z_0^I,\wt_0^I\bigr), \label{tvars6.2} \\
&\Psi_\ell^{IJ}  = \Delta \ut_\ell^{IJ} - \Pi P_{0,\ell}^{IJ}\bigl(y_0^{00},\yv^{KL}_{\ell+1},\del_K\utv^{00}_\ell,\utv^{00}_\ell,\lambda_0,\rhot_0,\rhotv_\ell,
z_0^K,\wt_0^K,\wtv^K_\ell\bigr)  - \notag \\
&\quad \ep^1 \Pi P_{1,\ell}^{IJ}\bigl(\ep,y_0^{ij},y_1^{KL},\yv_{\ell+2}^{KL},\utv^{ij}_\ell,\del_K\utv^{ij}_{\ell},\ut^{ij}_{\ell+1},\del_K \ut^{KL}_{\ell+1},
\del_M\del_N \ut^{KL}_\ell, \ut^{KL}_{\ell+2},\lambda_0,\rhot_0,\rhotv_\ell,z_0^K,\wt_0^K,\wtv^K_\ell \bigr), \label{tvars6.3}
}
\lalign{tvars6s}{
&\Upsilon^{00}_0 = \lambda_0 + \frac{\ep}{4 a_0} \ip{1}{F^0\bigl(\ep,y_0^{ij},y_1^{KL},\utv_1^{ij},\del_K\del_L \ut^{ij},\del_K\ut^{ij},\lambda_0,
\rhot_0,z_0^I,\wt_0^I\bigr)}\label{tvars6.4} \\
&\Upsilon^{0J}_0  = z^J_0 + \frac{\ep}{4 a_0}\ip{1}{F^J\bigl(\ep,y_0^{ij},y_1^{KL},\utv_1^{ij},\del_K\del_L \ut^{ij},\del_K\ut^{ij},\lambda_0,
\rhot_0,z_0^I,\wt_0^I\bigr)}, \label{tvars6.5} \\
&\Upsilon^{IJ}_{\ell+2}  = y^{IJ}_{\ell+2} +\frac{1}{a_0}\ip{1}{ P_{0,\ell}^{IJ}\bigl(y_0^{00},\yv^{KL}_{\ell+1},\del_K\utv^{00}_\ell,\utv^{00}_\ell,\lambda_0,\rhot_0,\rhotv_\ell,
z_0^K,\wt_0^K,\wtv^K_\ell\bigr)} - \notag \\
& \frac{\ep^1}{a_0}\ip{1}{ P_{1,\ell}^{IJ}\bigl(\ep,y_0^{ij},y_1^{KL},\yv_{\ell+2}^{KL},\utv^{ij}_\ell,\del_K\utv^{ij}_{\ell},\ut^{ij}_{\ell+1},\del_K \ut^{KL}_{\ell+1},
\del_M\del_N \ut^{KL}_\ell, \ut^{KL}_{\ell+2},\lambda_0,\rhot_0,\rhotv_\ell,z_0^K,\wt^K_0,\wtv^K_\ell \bigr)}, \label{tvars6.6}
}
\lalign{tvars6ss}{
&\Psi^{00}_{\ell+1} = \ut^{00}_{\ell+1} + \del_I\ut^{I0}_\ell + \sum_{p=0}^{\ell}\binom{\ell}{p}\left[\frac{3}{2}
\frac{d^{{p+1}}\ln(a)}{dt^{p+1}\;}
\Bigl|_{t=0} \,
\bigl(\ut^{00}_{\ell-p}+\delta_{p0}  y^{00}_0\bigr) \right. \notag \\
& \qquad \qquad \left. + \frac{1}{2} \frac{d^{{p+1}}a}{dt^{p+1}\;}
\Bigl|_{t=0}  \,\delta_{IJ}\bigl(y^{IJ}_{\ell-p}+\ep^2 \ut^{IJ}_{\ell-p}\bigr)\right], \label{tvars6.7}\\
&\Psi^{0J}_{\ell+1} = \ut^{0J}_{\ell+1} + \del_I \ut^{IJ}_\ell + \ep \sum_{p=0}^{\ell}\binom{s}{p}\frac{5}{2}\frac{d^{p+1}\ln(a)}{dt^{p+1}\;}
\Bigl|_{t=0}\, \bigl(\ut^{0J}_{\ell-p}+\delta_{p0} y^{0J}_0\bigr), \label{tvars6.8}
}
\lalign{tvars6sss}{
&\Omega^0_0 = \wt_0 -\ep Q^0_0(\ep,y_0^{KL},\ut^{jk}_0,z_0^K,\wt_0^J), \label{tvars6.9} \\
&\Omega^j_{\ell+1} = \wt_{\ell+1}^i - Q^i_{0,\ell}\bigl(\lambda_0,\del_x^{\alpha}\rhot_0,\del_x^{\alpha}\rhotv_\ell,z_0^I,
\del_x^\alpha \wt^j,\del_x^\alpha \wtv^j_\ell \bigr) \notag \\
& \qquad \qquad -
\ep Q^i_{1,\ell} \bigl(\ep,y_0^{ij},y_1^{IJ},\yv^{IJ}_{\ell+1},\utv^{ij}_{\ell+1},\del_K\utv^{jk}_\ell,
\lambda_0,\del_x^{\alpha} \rhot_0,\del_x^{\alpha}\rhotv_\ell,z_0^I,
\del_x^\alpha \wt^j,\del_K \utv^{00}_\ell\bigr),
\label{tvars6.10} \\
&\Omega_{\ell+1}  = \rhot_{\ell+1}^i - Q_{0,\ell}\bigl(\lambda_0,\del_x^{\alpha}\rhot_0,\del_x^{\alpha}\rhotv_\ell,z_0^I,
\del_x^\alpha \wt^j,\del_x^\alpha \wtv^j_\ell \bigr) \notag \\
& \qquad \qquad -
\ep Q_{1,\ell} \bigl(\ep,y_0^{ij},y_1^{IJ},\yv^{IJ}_{\ell+1},\utv^{ij}_{\ell+1},\del_K\utv^{jk}_\ell,
\lambda_0,\del_x^{\alpha} \rhot_0,\del_x^{\alpha}\rhotv_\ell,z_0^I,
\del_x^\alpha \wt^j\bigr).\label{tvars6.11}
}
Gathering all of the maps \eqref{tvars6.1}-\eqref{tvars6.11} together, we define
\leqn{Ximap1}{
\Xi_\ell = \bigl( \Psiv_{\ell+2}^{00},\Psiv_{\ell+2}^{0J},\Psiv_{\ell}^{IJ},\Omega^0_0,\Omegav^j_{\ell+1},\Omegav_{\ell+1},
\Upsilon^{00}_0,\Upsilon^{0J}_0,\Upsilonv_{\ell+2}^{IJ}\bigr)^T,
}
with
\lgath{tvars7}{
\Psiv_\ell^{ij} = (\Psi_0^{ij},\ldots,\Psi_{\ell}^{ij}), \quad
\Upsilonv^{IJ}_{\ell+2} = (\Upsilon^{IJ}_{2},\ldots,\Upsilon^{IJ}_{\ell+2}), \label{tvars7.1} \\
\Omegav^j_{\ell+1} = (\Omega^j_1,\ldots,\Omega^j_{\ell+1}), \AND
\Omegav_{\ell+1}  = (\Omega_1,\ldots,\Omega_{\ell+1}). \label{tvars7.2}
}
We also define
\alin{XYZ}{
\Xc^{s}_{R,\delta,\ell} &= B_{R}(\Sbb{4}) \times \Sbb{3} \times \Ho^{s+1-\ell}(\Sbb{3})\times \Ho^{s-\ell}(\Sbb{3})\times B_\delta (\Ho^s)\times B_{R}\bigl(\Ho^{s}(\Rbb^3)\bigr),  \\
\Yc^{s}_{R,\delta,\ell} &= B_{R}(\Ho^{s+2})\times \prod_{p=1}^{\ell+2} H^{s+2-p} \times B_{R}(\Ho^{s+2}(\Rbb^3))\times \prod_{p=1}^{\ell+2} H^{s+2-p}(\Rbb^3)
\times B_{R}(\Ho^{s+2}(\Sbb{3})), \\
& \quad \times \prod_{p=1}^\ell \Ho^{s+2-p}(\Sbb{3})
\times B_{R}(H^{s})  \times \prod_{p=1}^{\ell+2} H^{s-p}(\Rbb^4) \times
\prod_{p=1}^{\ell+2} H^{s-p} \times (-\delta,\delta) \times B_R(\Rbb^3) \times (\Sbb{3})^{\ell},
\intertext{and}
\Zc^{s}_\ell & = \Ho^{s}\times \prod_{p=1}^{\ell+2} H^{s+2-p} \times \Ho^{s}(\Rbb^3)\times \prod_{p=1}^{\ell+2} H^{s+2-p}(\Rbb^3) \\
& \text{\hspace{3.0cm}} \times \prod_{p=0}^\ell \Ho^{s-p}(\Sbb{3})
\times H^{s+1}  \times \prod_{p=1}^{\ell+1} H^{s-p}(\Rbb^4) \times
\prod_{p=1}^{\ell+1} H^{s-p} \times \Rbb \times \Rbb^3 \times (\Sbb{3})^{\ell}.
}

\begin{prop} \label{XipropA} \mnote{[XipropA]}
Suppose $\ell \in \Zbb_{\geq 0}$, $s > 3/2 + \ell$, $R>0$, $C_s$ is the constant from the inequality \eqref{BA}, $\delta = \mu_0/(2 C_s)$ and set
\eqn{XipropA.1}{
\theta_\ell = (y_0^{ij},y_1^{IJ},\ut^{IJ}_{\ell+1},\ut^{IJ}_{\ell+2},\rhot_0,\wt_0^I)^T, \AND
\eta_\ell = \bigl(\utv^{00}_{\ell+2},\utv^{0J}_{\ell+2},\utv^{IJ}_\ell,\wt^0_0,\wtv^j_{\ell+2},\rhotv_{\ell+2},
\lambda_0,z^I_0,\yv^{IJ}_{\ell+2}\bigr).
}
Then there exists an $\ep_0>0$ such that the map
\eqn{XipropA.3}{
(-\ep_0,\ep_0) \times \Xc^{s}_{R,\delta,\ell}\times \Yc^{s}_{R,\delta,\ell}  \ni (\ep,\theta_\ell,\eta_\ell) \longmapsto \Xi_\ell \in \Zc^{s}_\ell
}
is analytic.
\end{prop}
\begin{proof}
By definition,
\eqn{XipropA.4}{
\frac{1}{\rho} = \frac{1}{\mu_0 + \lambda_0 + \rhot_0} = \frac{1}{\mu_0}\frac{1}{ (1 + \bigl( \lambda_0 + \rhot_0)/\mu_0 \bigr)} .
}
Since the map $k(r)=1/(1+r/\mu_0)$ is in $C^\omega((-\mu_0,\mu_0),\Rbb)$, it follows from Proposition \ref{analprop} that the map
\eqn{XipropA.4}{
(-\delta,\delta)\times B_\delta (\Ho^s) \ni (\lambda_0,\rhot_0) \longmapsto \frac{1}{\rho} \in H^s
}
is well defined and analytic for $\delta = \mu_0/(2C_s)$. With this map well defined, we can recover $w^I$ from $z^I_0$ and $\wt_0^I$ by the defining relation
$w^I = (z^I_0+ \wt^I_0)/\rho$. The rest of the proof now follows from a straightforward application of Proposition \ref{analprop} and Lemmas \ref{facts} and \ref{mlem}.
\end{proof}

\subsect{idata}{Initialization to an arbitrary order}

We are now ready to prove that there exists a large class of initial data that can be initialized to an arbitrary order.
\begin{thm} \label{pdata} \mnote{[pdata]}
Suppose $\ell \in \Zbb_{\geq 0}$, $s > 3/2 + \ell$, $R>0$, and $\delta = \mu_0/(2C_s)$. Then for
any $\tilde{\theta}_\ell \in  \Xc^s_{R,\delta,\ell}$, there exists an open neighborhood $(-\ep_0,\ep_0)\times U_{\tilde{\theta}_\ell}\subset  \Rbb \times \Xc^s_{\Rt,\delta,\ell}$
of $(0,\tilde{\theta}_\ell)$ $(\ep_0>0, \Rt>R)$ and a map
\eqn{pdata1}{
\eta_\ell = \bigl(\utv^{00}_{\ell+2},\utv^{0J}_{\ell+2},\utv^{IJ}_\ell,\wt^0_0,\wtv^j_{\ell+2},\rhotv_{\ell+2},
\lambda_0,z^I_0,\yv^{IJ}_{\ell+2}\bigr) \in C^\omega\bigl( (-\ep_0,\ep_0)\times U_{\tilde{\theta}_\ell}, \Yc_{\Rt,\delta,\ell}\bigr)
}
that satisfies
\eqn{pdata2}{
\Xi_\ell(\ep,\theta_\ell,\eta_\ell(\ep,\theta_\ell)) = 0 \AND   \wt^0_0(0,\theta_\ell) = \lambda(0,\theta_\ell) = z^I_0(0,\theta_\ell)=0
}
for all $(\ep,\theta_\ell) \in (-\ep_0,\ep_0)\times U_{\tilde{\theta}_\ell}$.
\end{thm}
\begin{proof} We first establish that given a $\tilde{\theta}_\ell\in \Xc^{s}_{R,\delta,\ell}$, the equation $\Xi_\ell |_{\ep=0} = 0$
has a solution.
\begin{lem} \label{Xisol} \mnote{[Xisol]}
For any $\tilde{\theta}_\ell\in \Xc^{s}_{R,\delta,\ell}$, there exists a $\tilde{R} > 0$ and a
$\tilde{\eta}_{\ell}\in \Yc^{s}_{\tilde{R}/2,0,\ell}$ that satisfies
\eqn{Xisol.1}{
\Xi_\ell(0,\tilde{\theta}_\ell,\eta_{\ell}) = 0.
}
\end{lem}
\begin{proof}
To begin, we consider the fixed data
\eqn{fix}{
\tilde{\theta}_\ell = (y_0^{ij},y_1^{IJ},\ut^{IJ}_{\ell+1},\ut^{IJ}_{\ell+2},\rhot_0,\wt_0^I)^T \in  \Xc^{s}_{\infty,\delta,\ell}.
}
Next, we note that $\Xi|_{\ep=0} = 0$ reduces to (see \eqref{tvars6.1}-\eqref{tvars6.11})
\lalign{Xiepz}{
\lambda_0 & = 0, \label{Xiepz.1} \\
z^J_0 & = 0, \label{Xiepz.2} \\
\Delta\ut^{00}_0  & = 4a_0\rhot_0, \label{Xiepz.3} \\
\Delta \ut^{0J}_0  & = 4a_0\wt^J_0,  \label{Xiepz.4} \\
\Delta \ut_p^{IJ} & = \Pi\Pt_{0,p}^{IJ}\bigl((\rhot_0,\wt_0^K,y_0^{00},\yv^{KL}_{p+1},\utv^{00}_p,\rhotv_p,\wtv^K_p\bigr), \label{Xiepz.5} \\
y^{IJ}_{p+2} & = \frac{1}{a_0}\ip{1}{\Pt_{0,p}^{IJ}\bigl(\rhot_0,\wt_0^K,y_0^{00},\yv^{KL}_{p+1},\utv^{00}_p,\rhotv_p,\wtv^K_p\bigr)},  \label{Xiepz.6} \\
\ut^{00}_{p+1} &= - \del_I\ut^{I0}_p - \sum_{q=0}^{p}\binom{p}{q}\left[\frac{3}{2}
\frac{d^{{q+1}}\ln(a)}{dt^{q+1}\;}
\Bigl|_{t=0} \,
\bigl(\ut^{00}_{p-q}+\delta_{q0}y^{00}_0 \bigr) + \frac{1}{2} \frac{d^{{q+1}}a}{dt^{q+1}\;}
\Bigl|_{t=0}  \,\delta_{IJ} y^{IJ}_{p-q} \right], \label{Xiepz.7}\\
\ut^{0J}_{p+1} &= - \del_I \ut^{IJ}_p  \label{Xiepz.8}\\
\wt_0 & = 0, \label{Xiepz.9} \\
\wt_{p+1}^i & =  \Qt^i_{0,p}\bigl(\lambda_0,\rhot_0,\rhotv_p,z^I,\wt^0,\wt^I_0,\wtv^j_p,\utv^{00}_p\bigr),
\label{Xiepz.10} \\
\rhot_{p+1} & = \Qt_{0,p}\bigl(\lambda_0,\rhot_0,\rhotv_p,z^I,\wt^0,\wt^I_0,\wtv^j_p \bigr).
\label{Xiepz.11}
}
where the maps
\alin{Ximaps}{
& \Pt_{0,p} \: : \:  B_\delta(\Ho^s)\times \Ho^{s}(\Rbb^3)\times \Rbb  \times (\Sbb{3})^{p-1} \times \Ho^{s+2} \\
& \text{\hspace{1.0cm}} \times  \prod_{q=1}^{p} H^{s+2-q}  \times \prod_{q=1}^{p} H^{s-q}
 \times \prod_{q=1}^{p} H^{s-q}(\Rbb^3)
\longrightarrow H^{s-p}(\Sbb{3}) && (0\leq p \leq \ell), \\
& \Qt_{0,p}^i \: : \:  (-\delta,\delta)\times B_\delta (\Ho^s) \times \prod_{q=1}^p H^{s-q} \times \Rbb^3
\times H^{s}  \times \Ho^s(\Rbb^3) \\
& \text{\hspace{1.5cm}} \times  \prod_{q=1}^{p} H^{s-q}(\Rbb^4)
 \times \Ho^{s+2} \times \prod_{q=1}^{p} H^{s+2-q}(\Rbb^4)
\longrightarrow H^{s-p}(\Rbb^4) && (0\leq p \leq \ell+1), \\
& \Qt_{0,p} \: :\: (-\delta,\delta)\times B_\delta (\Ho^s) \times \prod_{q=1}^p H^{s-q} \times \Rbb^3
\times H^{s} \times \Ho^s(\Rbb^3)  \\
& \text{\hspace{3.0cm}} \times  \prod_{q=1}^{p} H^{s-q}(\Rbb^4)
\longrightarrow H^{s-p-1} && (0\leq p \leq \ell+1)
}
are all analytic for $\delta = \mu_0/(2C_s)$.

The  invertibility of the Laplacian $\Delta : \Ho^{k+2} \rightarrow \Ho^{k}$ then implies, by equations \eqref{Xiepz.3}-\eqref{Xiepz.4},
that
\leqn{solve1}{
\ut_0^{00} = 4 a_0 \Delta^{-1}\rhot_0 \AND \ut_0^{0J} = 4a_0 \Delta^{-1} \wt_0^J.
}
Substituting these into \eqref{Xiepz.5}, \eqref{Xiepz.6}, \eqref{Xiepz.7}, \eqref{Xiepz.10}, and
\eqref{Xiepz.11} (for $(p=0)$) gives
\lalign{solve2}{
\ut_0^{IJ} &= \Delta^{-1}\Pi\Pt_{0,0}^{IJ}\bigl(\rhot_0,\wt_0^K,y_0^{00},\ut_0^{00}\bigr), \label{solve2.1} \\
y_2^{IJ} &=  \frac{1}{a_0}\ip{1}{\Pt_{0,0}^{IJ}\bigl(\rhot_0,\wt_0^K,y_0^{00},\ut_0^{00}\bigr)} ,\label{solve2.2} \\
\ut^{00}_1 & = -\del_I \ut^{0I}_0 - \frac{a'(0)}{2a(0)}\bigl(3\ut^{00}_0 + 3y^{00}_0 + a(0) \delta_{IJ} y^{IJ}_0\bigr),
\label{solve2.3} \\
\wt_{1}^i & =  \Qt^i_{0,0}\bigl(\lambda_0,\rhot_0,z^I_0,\wt^j_0,\ut^{00}\bigr),
\label{solve2.4}
\intertext{and}
\rhot_{1} & = \Qt_{0,0}\bigl(\lambda_0,\rhot_0,z^I_0,\wt^j_0\bigr).
\label{solve2.5}
}
From \eqref{Xiepz.7} and \eqref{solve2.1}, we then obtain
\leqn{solve3}{
\ut^{0J}_1 = -\del_{I}\ut^{IJ}_0.
}
Substituting \eqref{solve1}-\eqref{solve3} into \eqref{Xiepz.5}, we find
\leqn{solve4}{
\ut^{IJ}_1 = \Delta^{-1}\Pi \bigl(\rhot_0,\wt_0^K,y_0^{00},y^{KL}_{2},\utv^{00}_1,\rhot_1,\wt^K_1\bigr).
}

With the base case covered, we proceed by induction. So assume that $\{\ut^{ij}_q,y^{IJ}_{q+1},\rhot_{q},\wt^j_q\}_{q=1}^p$
solves \eqref{Xiepz.5}-\eqref{Xiepz.8} and \eqref{Xiepz.10}-\eqref{Xiepz.11} for $0\leq q \leq p \leq \ell-1$. Then
it is clear that we can immediately use \eqref{Xiepz.6} - \eqref{Xiepz.8}, and  \eqref{Xiepz.10}-\eqref{Xiepz.11} to
determine $y^{IJ}_{p+2}$, $\ut^{0j}_{p+1}$, $\wt^j_{p+1}$, and $\rhot_{p+1}$ from  $\{\ut^{ij}_q,y^{IJ}_{q+1},\rhot_{q},\wt^j_q\}_{q=1}^p$.
We then substitute these into \eqref{Xiepz.5} to determine $\ut^{IJ}_{p+1}$ which completes the induction step.
With the  $\{\ut^{ij}_q,y^{IJ}_{q+1},\rhot_{q},\wt^j_q\}_{q=1}^\ell$ determined, similar
arguments show that we can use these along with the initial data $\{\ut^{IJ}_{\ell+1},\ut^{IJ}_{\ell+2}\}$
to find $\{\ut^{0j}_{\ell+1},\ut^{0j}_{\ell+2},\rhot_{\ell+1},\rhot_{\ell+2},\wt^j_{\ell+1},\wt^j_{\ell+2}\}$.
\end{proof}

Having constructed a solution to $\Xi_\ell|_{\ep=0} =0$, we will use the implicit function
theorem to find solutions to $\Xi_\ell =0$ for $\ep > 0$. However, to apply the implicit function
theorem, we must first establish that the partial derivative of $\Xi_\ell$ with
respect to $\eta_\ell$ is an isomorphism.
\begin{lem} \label{DXi} \mnote{[DXi]}
Suppose $\theta_\ell \in \Xc^{s}_{\infty,\delta,\ell}$,
$\eta_\ell \in \Yc^{s}_{\infty,\delta,\ell}$, and let $\Xi_{\theta_\ell}$ be the map
defined by $\Xi_{\theta_\ell}(\cdot) = \Xi(0,\theta_\ell,\cdot)$. Then
the derivative
\eqn{DXi.1}{
D\Xi_{\theta_\ell}(\eta)\; : \;  \Yc^{s}_{\infty,\infty,\ell} \longrightarrow \Zc^{s}_\ell
}
is a linear isomorphism.
\end{lem}
\begin{proof}
Fix
\eqn{DXi.2}{\sigma_\ell = \bigl(\psiv_{\ell+2}^{00},\psiv^{0J}_{\ell+2},\psiv^{IJ}_\ell,\omega^0_0,\omegav^{j}_{\ell+1},\omegav_{\ell+1}, \upsilon^{00}_0,
\upsilon^{0J}_{0},\upsilonv^{IJ}_{\ell+2} \bigr) \in \Zc^s_\ell,
}
and let
\eqn{DXi.3}{
\delta\eta_\ell =  \bigl(\delta\utv^{00}_{\ell+2},\delta\utv^{0J}_{\ell+2},
\delta\utv^{IJ}_\ell, \delta\wt^0, \delta \wtv^j_{\ell+1}, \delta\rhotv_{\ell+1},
\delta \lambda_0, \delta z^I_0,\delta \yv^{IJ}_{\ell+2}\bigr).
}
Then from \eqref{tvars6.1}-\eqref{tvars6.11}, it is not difficult to see that
the equation
\eqn{DXi.4}{
D\Xi_{\theta_\ell}(\eta_\ell)\cdot \delta\eta_\ell = \sigma_\ell
}
is equivalent to the following system:
\alin{DXi.5}{
& \delta \lambda_0  = \upsilon^{00}_0, \\
& \delta z^J_0  = \upsilon_0^{0J}, \\
&\Delta \delta \ut^{00}_0   = \psi^{00}_0, \\
&\Delta \ut^{0J}_0  = \psi^{00}_0,   \\
&\Delta \ut_p^{IJ} - \Pi \Lambda_{p}^{IJ}\bigl(\delta \yv^{KL}_{p+1},\delta \utv^{00}_p,\delta\rhotv_p,\delta\wtv^K_p\bigr)
= \psi_p^{IJ},  \\
& y^{IJ}_{p+2} - \ip{1}{\Lambda_{p}^{IJ}\bigl(\delta \yv^{KL}_{p+1},\delta \utv^{00}_p,\delta \rhotv_p,\delta \wtv^K_p\bigr)}
= \upsilon^{IJ}_{p+2},   \\
&\delta \ut^{00}_{p+1} + \del_I \delta \ut^{I0}_p + \sum_{q=0}^{p}\binom{p}{q}\left[\frac{3}{2}
\frac{d^{{q+1}}\ln(a)}{dt^{q+1}\;}\Bigl|_{t=0} \,\delta\ut^{00}_{p-q} + \frac{1}{2} \frac{d^{{q+1}}a}{dt^{q+1}\;}
\Bigl|_{t=0}  \,\delta_{IJ} \delta y^{IJ}_{p-q} \right] = \psi^{00}_{p+1},\\
&\delta \ut^{0J}_{p+1} + \del_I \delta \ut^{IJ}_p = \psi^{0J}_{p+1} \\
& \delta \wt_0  = \omega^0_0, \\
& \delta \wt_{p+1}^i - \Theta^i_{p}\bigl(\delta \rhotv_p,\delta\wt_0^0,\delta \wtv^j_p,\delta \utv^{00}_p\bigr) = \omega^{i}_{p+1},
\intertext{and}
& \delta \rhot_{p+1} - \Theta_{p}\bigl( \delta \rhotv_p,\delta\wt_0^0,\delta \wtv^I_p \bigr) = \omega_{p+1},
}
where $\Lambda^{IJ}_p$, $\Theta^i_p$, and $\Theta_p$ are linear maps that depend implicitly on
$\eta_\ell$. This system has the same structure as the system \eqref{Xiepz.1}-\eqref{Xiepz.11} from Lemma \ref{Xisol}, and
a slight variation of the arguments used in the proof of the Lemma can be used to establish
the existence of
a unique solution for the given $\sigma_\ell$. In particular, this shows that $D\Xi_{\theta_\ell}(\eta_\ell)$ is an isomorphism.
\end{proof}

By Proposition \ref{XipropA}, the map
$\Xi_\ell \: : \: (-\ep_0,\ep_0) \times \Xc^{s}_{R,\delta,\ell}\times \Yc^{s}_{R,\delta,\ell} \rightarrow \Zc^{s}_\ell$ is
well defined and analytic for $\ep_0>0$ small enough. Lemmas \ref{Xisol} and \ref{DXi} then allow us to apply an analytic
version of the implicit function theorem (see \cite{Deim}, Theorem 15.3) to conclude the existence of (shrinking $\ep_0$ if necessary)
an open neighborhood $(-\ep_0,\ep_0)\times U_{\tilde{\theta}_\ell}\subset  \Rbb \times \Xc_{\Rt,\delta,\ell}$
of $(0,\tilde{\theta}_\ell)$  and an analytic map
$\eta_\ell \: : \: (-\ep_0,\ep_0)\times U_{\tilde{\theta}_\ell} \rightarrow \Yc_{\Rt,\delta,\ell}$
that satisfies
$\Xi_\ell(\ep,\theta_\ell,\eta_\ell(\ep,\theta_\ell)) = 0$
for all $(\ep,\theta_\ell) \in (-\ep_0,\ep_0)\times U_{\tilde{\theta}_\ell}$.
\end{proof}

\sect{pnexp}{Post-Newtonian expansions}

\subsect{lim}{The limit equation}

Before discussing the $\ep \searrow 0$ limit equation for the system \eqref{nonloc1},
we first consider local existence and uniqueness of solutions to the cosmological Poisson-Euler
equations.
\begin{prop} \label{limA} \mnote{[limA]}
Suppose $s > 3/2+ 3 + \ell$, $\delta = \mu_0/(2C_s)$,
$\rhot_0 \in B_\delta(\Ho^s)$, and  $\wt^I_0\in \Ho^{s}(\Rbb^3)$. Then there exists a maximal time $T_0$ and
a unique
solution
\eqn{limA1}{
\oset{0}{\rho}(t) \in X_{T_0,\ell+2,s}, \quad \oset{0}{w}{}^I(t) \in X_{T_0,\ell+2,s}(\Rbb^3), \quad  \oset{0}{\Phi}(t)\in
X_{T_0,\ell+4,s+2},
}
to the Poisson-Euler \eqref{PE.1}-\eqref{PE.3} with initial data $\oset{0}{\rho}(0) = \mu_0+\rhot$ and
$\oset{0}{w}{}^I = \wt^I/(\mu_0+\rhot)$. Moreover, this solution satisfies
\eqn{limA2}{
\ip{1}{\oset{0}{\rho}(t)\oset{0}{w}{}^I(t) }_{L^2} = \ip{\Pi \oset{0}{\rho}(t)}{\del_I\oset{0}{\Phi}(t)}_{L^2} = 0
}
for all $t\in [0,T_0)$.
\end{prop}
\begin{proof}
As the system Poisson-Euler \eqref{PE.1}-\eqref{PE.3} is clearly a (nonlocal) symmetric hyperbolic system, the
statements concerning existence and uniqueness follow immediately from standard theory. To prove the second
statement, we observe that $\oset{0}{\rho} \oset{0}{w}{}^I$ satisfies
\eqn{limA3}{
\del_t(\oset{0}{\rho}\oset{0}{w}{}^J) = -\del_I(\oset{0}{w}{}^I\oset{0}{\rho}\oset{0}{w}{}^J)
-\frac{1}{a(t)}\left( \del^J f(\oset{0}{\rho}) + \oset{0}{\rho} \del^J \oset{0}{\Phi} + \frac{3a'(t)}{2}
\oset{0}{\rho}\oset{0}{w}{}^J\right).
}
Taking the $L^2$ inner product of this equation with 1 yields
\leqn{limA4}{
\del_t\ip{1}{\oset{0}{\rho}\oset{0}{w}{}^J}_{L^2} =
-\frac{1}{a(t)}\left( \ip{\Pi \oset{0}{\rho} }{ \del^J \oset{0}{\Phi}}_{L^2} + \frac{3a'(t)}{2}
\ip{1}{\oset{0}{\rho}\oset{0}{w}{}^J}_{L^2}\right).
}
By \eqref{PE.3}, we have
$a(t)\ip{\Pi \oset{0}{\rho} }{ \del^J \oset{0}{\Phi}}_{L^2}
= \ip{\Delta \oset{0}{\Phi}}{\del^J\oset{0}{\Phi}}_{L^2}$
and hence
\eqn{limA6}{
a(t)\ip{\Pi \oset{0}{\rho} }{ \del^J \oset{0}{\Phi}}_{L^2}
=  -\int_{\Tbb^3} \del_{J}\del_I\oset{0}{\Phi}\, \del^I \oset{0}{\Phi} \, d^3 x
= -\frac{1}{2}  \int_{\Tbb^3} \del_J\bigl(\del_I\oset{0}{\Phi}\del^I \oset{0}{\Phi}\bigr) \, d^3 x = 0.
}
Substituting this into \eqref{limA4} gives
\leqn{limA7}{
\del_t \ip{1}{\oset{0}{\rho}\oset{0}{w}{}^J}_{L^2} =
-\frac{3a'(t)}{2 a(t)} \ip{1}{\oset{0}{\rho}\oset{0}{w}{}^J}_{L^2}.
}
By assumption, $\ip{1}{\bigl(\oset{0}{\rho}\oset{0}{w}\bigr)|_{t=0}}_{L^2} = \ip{1}{\wt_0^I}_{L^2} = 0$ which combined with the
differential equation \eqref{limA7} shows that $\ip{1}{\oset{0}{\rho}(t)\oset{0}{w}(t)}_{L^2}=0$ for all $t\in [0,T_0)$.
\end{proof}

From \cite{Scho88}, we know that the appropriate $\ep \searrow 0$ limit equation for the system \eqref{nonloc1}
is
\lalign{limeqn}{
A^0_0 \del_t W &=  A^I_0 \del_I W +  F_0 + C^I\del_I\omega, \label{limeqn.1}\\
C^I\del_I W  & = 0 \label{limeqn.2},
}
where
\leqn{limeqn1}{
A^0_0 = \begin{pmatrix} A^0_{G,0} & 0 \\ 0 & A_{M,0} \end{pmatrix}, \AND
A^0_{G,0} = \begin{pmatrix} a(t) & 0 & 0 & 0 \\ 0 & \delta^{IJ} & 0 & 0 \\ 0 & 0 & 1 & 0 \\ 0 & 0 & 0 & 1 \end{pmatrix}.
}
The relationship between the Poisson-Euler equations \eqref{PE.1}-\eqref{PE.3} and the limit
equation \eqref{limeqn.1}-\eqref{limeqn.2} is given by the following Proposition. Here and
in the following section, we require the following evolution spaces
\eqn{Xcspace}{
\Xc_{T,\ell,s} = \bigcap_{p=0}^{\ell+1} C^p\bigl([0,T), \Hc^{s-p}\bigr).
}

\begin{prop} \label{limB} \mnote{[limB]}
Suppose $s > 3/2+ 3 + \ell$, $\delta = \mu_0/(2C_s)$,
$\rhot_0 \in B_\delta(\Ho^s)$, $\wt^I_0\in \Ho^{s}(\Rbb^3)$, and let $\{\oset{0}{\rho},\oset{0}{w}{}^I,\oset{0}{\Phi}\}$ and
$T_0$ be as in proposition \ref{limA}. Then
\eqn{limB1}{
\oset{0}{W} = \bigl(0,0,0,0,\oset{0}{\rho},(0,\oset{0}{w}{}^I)\bigr) \in \Xc_{T_0,\ell+2,s} \AND \omega = \bigl(\delta^i_0\delta^j_0 \del_t\oset{0}{\Phi},0,0,0,0,0) \in \Xc_{T_0,\ell+3,s+1}
}
solve the limit equation \eqref{limeqn.1}-\eqref{limeqn.2}.
\end{prop}
\begin{proof}
The proof follows directly from substituting $\oset{0}{W}$ and $\omega$ into  \eqref{limeqn.1}-\eqref{limeqn.2} while using
\eqref{PE.1}-\eqref{PE.3}, \eqref{AFdec2.1}-\eqref{AFdec2.3}, \eqref{Adef.3}-\eqref{Adef.4}, and \eqref{Fdef.1}-\eqref{Fdef.2}.
\end{proof}

\subsect{proof}{Proof of Theorem \ref{mthm}}

We are now ready to prove Theorem \ref{mthm} and thus establish the existence of cosmological post-Newtonian expansions to
arbitrary order.

\begin{proof}[Proof of Theorem \ref{mthm}]
Given $y_0^{ij} \in \Sbb{4}$, $y_1^{IJ}\in \Sbb{3}$,
$\rhot_0 \in B_R(\Ho^s)$, $\wt^I_0\in \Ho^{s}(\Rbb^3)$, $\ut^{IJ}_\ell \in \Ho^{s+1-\ell}(\Sbb{3})$,
and $\ut^{IJ}_{\ell+1} \in \Ho^{s-\ell}(\Sbb{3})$, let
\gath{pmthm1}{
\delt{p} \ub^{IJ}_\ep = y_p^{IJ} + \ep^2 \ut_p^{IJ}(\ep), \quad
\delt{p} \ub^{0J}_\ep  = \ep(\delta_p^0 y^{0J}_0 + \ut_p^{0J}(\ep)), \quad
\delt{p} \ub^{00}_\ep  = \delta_p^0 y^{00}_0 + \ut_p^{00}(\ep), \quad p=0,1 \\
\rho_\ep|_{t=0} = \mu_0+\lambda_0(\ep)+ \rhot_0, \AND
(\rho w^I)|_{t=0} =  \frac{z^I_0(\ep) + \wt_0^I}{\mu_0+\lambda_0(\ep)+ \rhot_0}
}
be the initial data from Theorem \ref{pdata}. By construction, this data solves the constraint equations
\eqref{ceqns.1}-\eqref{ceqns.3}, depends analytically on $\ep$, and satisfies \eqref{tvars1.1}-\eqref{tvars1.3}.
In particular, this implies by \eqref{Wdef1}, \eqref{Phidef}, and Lemma \ref{philem} that
\leqn{pmthm1}{
W_\ep(t) = \bigl(u^{ij}_{0,\ep}(t), W^{ij}_{I,\ep}(t),u^{ij}_{\ep}(t),\phi^{ij}_{\ep}(t),\wv_\ep(t)\bigl)^T
}
satisfies
\leqn{pmthm2}{
W_\ep(0) \in C^\omega((-\ep_0,\ep_0),\Hc^s),
}
and
\leqn{pmthm4}{
\bnorm{\delt{p} W_\ep}_{\Hc^{s-p}} \lesssim 1 \quad \text{for $p=1,2,\ldots,\ell+1$}.
}
Together, Proposition \ref{limB} and \eqref{pmthm1}-\eqref{pmthm2} allow us to apply to Theorem 3 of \cite{Scho88}
and conclude (shrinking $\ep_0$ if necessary) that for any $T<T_0$ there exists maps
\alin{pmthm5}{
&W_\ep \in \Xc_{T,\ell+2,s} \qquad 0<\ep < \ep_0, \\
&\oset{p}{W} \in \Xc_{T,\ell+2-p,s-p} \qquad p=1,2,\ldots,\ell, \\
&\oset{p}{W}_\ep \in \Xc_{T,1,s-\ell-1} \qquad (p,\ep)\in \Zbb_{\geq \ell+1}\times (0,\ep_0),
}
such that
\begin{itemize}
\item[(i)] $W_\ep (t,x^I)$ solves equation \eqref{nonloc1} on the spacetime region $(t=x^0,x^I)\in M=[0,T)\times \Tbb^3$,
\item[(ii)] $\oset{p}{W}$  $(1\leq p \leq \ell)$ satisfies a linear (nonlocal) symmetric hyperbolic system
that depends only on $\{\: \oset{q}{W} \: | \: q=0,1,\ldots,p-1 \: \}$ where $\oset{0}{W}$ is a defined in Proposition
\ref{limB},
\item[(iii)]  $\oset{p}{W}_\ep$  $(p \geq \ell+1)$ and $W_\ep$ satisfy the estimates
\eqn{pmthm6}{
\norm{W_\ep(t)}_{\Hc^s} + \ep \norm{\del_t W_\ep (t)}_{\Hc^{s-1}} \lesssim 1, \AND
\norm{\oset{q}{W}_\ep (t)}_{\Hc^{s-\ell-1}} + \ep \norm{\del_t \oset{q}{W}_\ep (t)}_{\Hc^{s-\ell-2}} \lesssim 1
}
for all $(t,\ep) \in [0,T)\times (0,\ep_0)$, and
\item[(iv)] $W_\ep$ admits a convergent expansion (uniform for $0<\ep<\ep_0$) of the form
\eqn{pmthm7}{
W_\ep = \oset{0}{W} + \sum_{p=1}^\ell \ep^p\, \oset{p}{W} + \sum_{p=\ell+1}^\infty \, \oset{p}{W}_\ep
}
where the expansion is convergent in $C^0([0,T),\Hc^{s-\ell-2})$.
\end{itemize}
Finally, similar arguments as in the proof of Proposition 6.1 in \cite{Oli06} can be used to show that
$\{\ub^{ij}_\ep = \ep^{-1} u^{ij}_\ep,\rho_\ep,w^i_\ep\}$ determine, via formulas \eqref{metrecA}-\eqref{wdef.intro}, a solution to the
Einstein-Euler equations in the harmonic gauge and moreover that  $\del_t \ub^{ij}_\ep = \ep^{-1} u_{0,\ep}^{ij}$
and $\del_I \ub_\ep^{ij} = W_{I,\ep}^I+\del_I\Phi_\ep^{ij}$. This combined with the statements (i)-(iv) above
completes the proof.
\end{proof}

\bigskip

\emph{Acknowledgements}
Part of the research for this article was completed while I was visiting the Mittag-Leffler Institute
during the \emph{Geometric, Analysis, and General Relativity program} in the Fall
of 2008. I thank the institute for its support and hospitality. I would also like to thank the referees
for their comments and criticisms which helped to improve the exposition of this article.  



\end{document}